\documentclass[preprint]{aastexr}

\shorttitle{Solar Observing with ALMA}
\shortauthors{Bastian et al.}

\begin{document}

\title[Solar Observing with ALMA]{Solar Observing with the Atacama Large Millimeter-Submillimeter Array\footnote{Accepted for publication in Frontiers in Astronomy and Space Science}} 

\correspondingauthor{T. S. Bastian}
\email{tbastian@nrao.edu}

\author{T. S. Bastian}
\affiliation{National Radio Astronomy Observatory, 520 Edgemont Road, 
Charlottesville, VA 22903 USA}

\author{M. Shimojo}
\affiliation{National Astronomical Observatory of Japan, 2-21-1, Osawa, Mitaka, Tokyo 181-8588, Japan}

\author{M. B\'{a}rta}
\affiliation{Astronomical Institute, Academy of Sciences, Fri\v{c}ova 298, 251 65 Ond\v{r}ejov, Czech Republic}

\author{S. M. White}
\affiliation{pace Vehicles Directorate, Air Force Research Laboratory, 3550 Aberdeen Avenue SE, Kirtland AFB, NM 87117-5776, USA}

\author{K. Iwai}
\affiliation{Institute for Space-Earth Environmental Research, Nagoya University, Furo, Chikusa, Nagoya, 464-8601, Japan}

\begin{abstract}
\noindent The Atacama Large Millimeter-submillimeter Array (ALMA), sited on the high desert plains of Chajnantor in Chile, has opened a new window onto solar physics in 2016 by providing continuum observations at millimeter and sub-millimeter wavelengths with an angular resolution comparable to that available at optical (O), ultraviolet (UV), extreme ultraviolet (EUV), and X-ray wavelengths, and with superior time resolution. In the intervening years, progress has been made testing and commissioning new observing modes and capabilities, in developing data calibration strategies, and in data imaging and restoration techniques. Here we review ALMA current solar observing capabilities, the process by which a user may propose to use the instrument, and summarize the observing process and work flow. We then discuss some of the challenges users may encounter in imaging and analyzing their data.  We conclude with a discussion of additional solar observing capabilities and modes under consideration that are intended to further exploit the unique spectral coverage provided by ALMA.
\end{abstract}


\section{Introduction}

The Atacama Large Millimeter-submillimeter Array (ALMA) is a high-performance, general-purpose telescope that has opened a new frontier of astrophysics, offering fundamentally new observations at mm-$\lambda$ and submm-$\lambda$. ALMA first became available to the wider scientific community more than a decade ago, at the end of 2011 (ALMA observing Cycle 0) but it was not until five years later in late 2016 that solar observations first became available with ALMA (observing Cycle 4), Since then, the solar physics community has embarked on a number of ambitious solar observing programs, providing new insights into outstanding problems in solar physics. Results from several of these programs are presented in the Frontiers collection of articles and reviews entitled ``The Sun Seen with the Atacama Large mm and sub-mm Array (ALMA) - First Results ".

Solar observing with ALMA is not without challenges. First, a number of technical barriers needed to be overcome to enable solar observing – and continue to be addressed as new observing modes and capabilities are considered. Second, as an interferometer the imaging and data reduction techniques employed by ALMA are not necessarily familiar to the wider solar community and a learning curve must be surmounted. In fact, solar imaging and reduction techniques remain under development in some cases. Hence, the ALMA solar user community is small, but growing. Finally, as a new window onto the Sun, the observations themselves pose interesting challenges to our understanding of physical processes on the Sun! 

In the remainder of this section we introduce the key attributes of ALMA and summarize the basic elements of the technical and operational approach to observing the Sun. Readers may wish to consult additional resources: a general overview of the instrument may be found in \citet{Wootten2009} while details regarding solar observing and data calibration are presented by \citet{Shimojo2017} and \citet{White2017}. In \S2 we summarize ALMA’s current observing capabilities, with an emphasis on new capabilities that have been commissioned since Cycle 4. In \S3 we discuss the process by which an interested scientist may propose to observe with ALMA. We then describe, for successful observing programs, the observing process and the subsequent work flow leading to data delivery to the observer. In \S4 we discuss some of the lessons learned from experience with ALMA imaging and data analysis. In \S5 we discuss future observing modes and capabilities that will enhance solar science with ALMA. We conclude in \S6.

\subsection{Organization of the Observatory}

Before describing the instrument itself we briefly describe the organization of the observatory. The observatory is fundamentally international in nature, a partnership between the United States National Science Foundation, the European Organization for Astronomical research in the Southern Hemisphere, and the National Institutes of Natural  Sciences of Japan, in cooperation with the Republic of Chile. Management of the of the observatory is consolidated under the Joint ALMA Observatory (JAO\footnote{See https://www.almaobservatory.org}). The JAO is based in Santiago, Chile, and is responsible for operating, maintaining, enhancing, and optimizing ALMA on behalf of the wider scientific community. The JAO organizes and oversees calls for proposals and their review, executes approved observing programs, and archives the data. The telescope itself is sited approximately 1500 km north of Santiago at the Array Operations Site (AOS). Telescope operations are supported and maintained from the Operations Support Facility (OSF) at a distance of about 30~km from the AOS. The OSF also supports assembly, integration, and verification of technology before it is moved to the AOS. 

The ALMA scientific community is supported by ALMA Regional Centers (ARCs) affiliated with each ALMA partner\footnote{The roles and responsibilities of the ARCs can be found at https://almaobservatory.org/en/about-alma/global-collaboration}. These are operated by  the National Radio Astronomy Observatory (NRAO) on behalf of North America, the European Southern Observatory (ESO) on behalf of its member states, and the National Astronomical Observatory of Japan (NAOJ) on behalf of East Asia. The North American and East Asia ARCs are co-located with their national observatories whereas the European ARC follows a distributed model of seven ARC nodes coordinated from ESO headquarters in Garching. The ARC node with responsibilities for the European solar community is based in Ondrejov, Czech Republic. The ARCs are each advised by scientific advisory committees to ensure that user concerns and priorities are known and addressed. The overall observatory is advised by the ALMA Scientific Advisory Committee. 

The role of the ARCs or ARC nodes in the workflow of a given project is discussed in greater detail in \S3. It is important to note that the ARCs also play an important role in education and outreach. For example, the ARCs are used to organize ALMA Community Workshops prior to each call for proposal. The goal of those meetings is to help the local community with proposal preparation. 

\subsection{Overview of the Instrument}

ALMA is located on the Chajnantor plateau of the Chilean Andes at a latitude of $-23^\circ$, a longitude of $-67.8^\circ$, and an elevation of over 5000~m. It is a general-purpose telescope designed to operate at mm-$\lambda$ and submm-$\lambda$ in order to address an extremely broad program of astrophysics – from cosmology, to star and planet formation, to astrochemistry. ALMA is an interferometer comprising an array of antennas that sample the Fourier transform of the brightness distribution within the field of view. A given pair of antennas measures a single Fourier component, an amplitude and a phase referred to as a complex {\sl visibility}, corresponding to a spatial frequency defined by the antenna spacing and orientation. A given antenna spacing is referred to as an antenna {\sl baseline}. Long antenna baselines measure small angular scales and short baselines measure large angular scales in the brightness distribution. The Fourier domain in which the measurements are made is referred to as the aperture plane or the {\sl uv} plane, where {\sl u} and {\sl v} refer to the coordinates of antenna baselines, typically measured in wavelength units. An array of $N$ antennas has $N(N-1)/2$ independent baselines. The instantaneous {\sl uv} sampling provided by an array is sometimes referred to as the “snapshot” {\sl uv} coverage. For sources that are static in time, one can exploit the fact that the array geometry, as viewed from the source, changes due to Earth’s rotation, allowing the {\sl uv} plane sampling to be filled in with time. This technique is referred to as Earth rotation aperture synthesis. 

\begin{figure}[h!]
\begin{center}
\includegraphics[width=12cm]{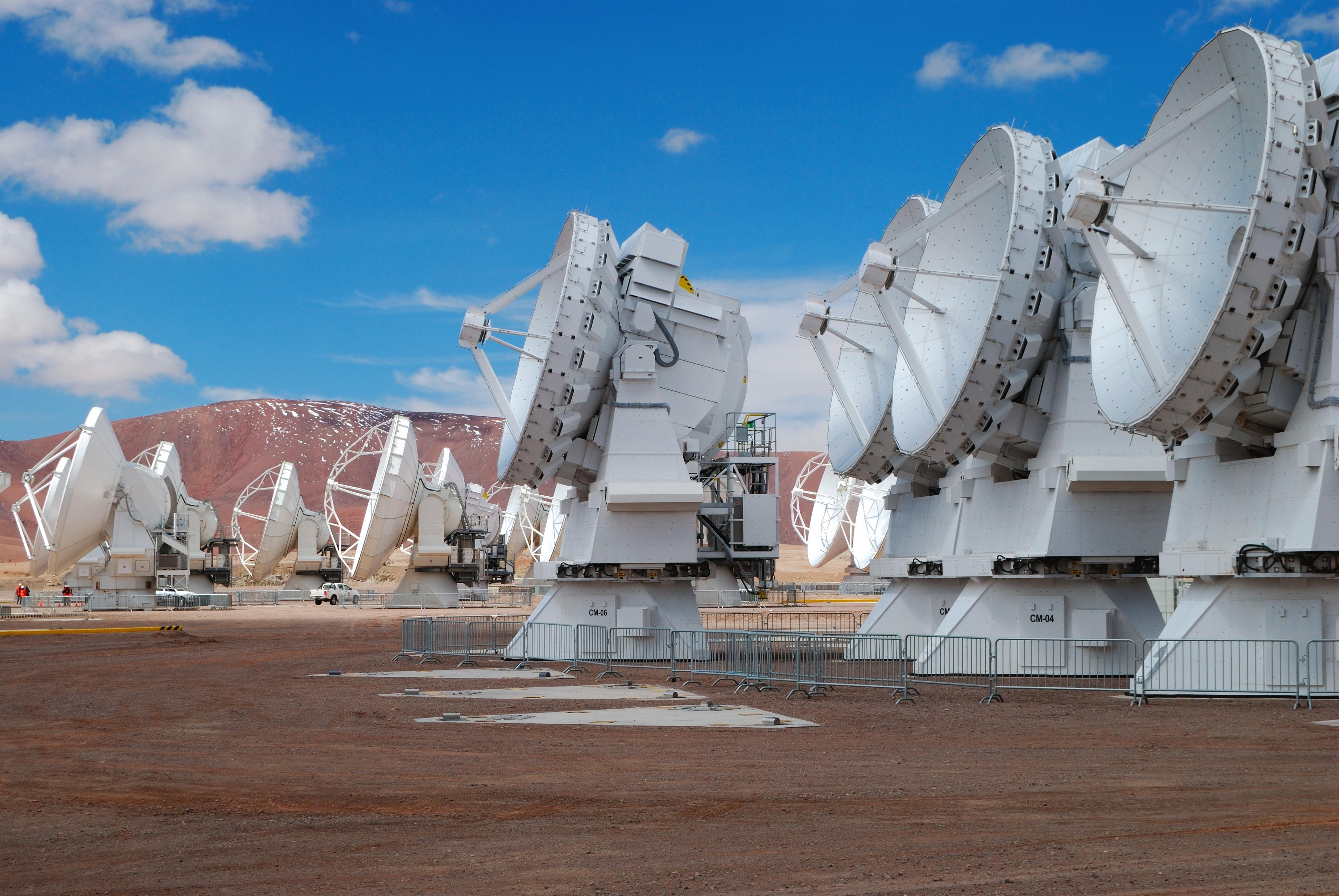}
\end{center}
\caption{ALMA is an array composed of 50 x 12~m movable antennas (left background,), $12 \times 7$~m fixed antennas (right foreground), and $4 \times 12$~m total power antennas. }\label{fig:1}
\end{figure}

ALMA has a total of 66 high performance antennas. The 12-m array is composed of $50 \times 12$ m antennas and is reconfigurable. That is, the antennas may be moved to change the distribution of baselines and, hence, sampling in the {\sl uv} plane in order to change the angular resolution and surface brightness sensitivity of the instrument. Baseline lengths can be as small as 15~m and as large as 16~km although for solar observing, array configurations are currently restricted to the four most compact configurations of the ten array configurations in general use (see Table 2). With all 50 antennas operative in the 12-m array, a total of 1225 independent baselines are available. The field of view (FOV) of the 12-m array is determined by the response of a single antenna which, to first order may be described by a modified Airy function with a main lobe and sidelobes. The main lobe is well described by a Gaussian. ALMA antennas are of Cassegrain design; accounting for the taper of the antenna illumination pattern and blockage by the subreflector and the quadrupod support structure, the HPBW of the Gaussian main beam – also referred to as the primary beam – is given by $\theta_{FOV} = 1.13\lambda/D =19.42” \lambda_{mm}$.  The Atacama Compact Array (ACA) comprises $12 \times 7$ m antennas and four 12~m total power antennas. The 7~m antennas are fixed and provide lower-resolution imaging to complement that provided by the 12-m array, or it can be used in a standalone mode. In the case of solar observations, the array of 7~m antennas is used to supplement the angular coverage provided by the 12-m array as discussed further in \S2. The FOV of a 7~m antenna is $\theta_{FOV} = 33.3” \lambda_{mm}$. The four 12~m antennas are available for making total power measurements. These play an important role in solar observing by effectively filling in the smallest spatial frequencies that are not otherwise measured by the 12-m array and the 7~m array of antennas as we discuss further in \S2.5 and \S4.4. 

Interferometric (INT) observations made by with the 12-m array use the Baseline Correlator (BLC) to produce visibility data. The ACA correlator may be used to correlate antennas in the 7-m array. Alternatively, as is the case for solar observations (see \S1.2.3 below), correlations between antennas in the ACA and the 12-m array may be processed together through the BLC. Continuum observations are performed using the ALMA Time Domain Mode (TDM). A given target is observed in two frequency ranges above and below the local oscillator (LO) frequency – the upper and lower sidebands. Each side band is subdivided into two spectral windows, each of 2 GHz bandwidth. Hence, a total of four spectral windows is observed with a combined bandwidth of 8 GHz. Available observing bands and spectral windows are summarized in Table~1. Each spectral window is currently fixed in frequency as a result of the gain reduction approach adopted (see \S1.3.1). The TDM mode coarsely channelizes each spectral window into 128 frequency channels. During calibration and data reduction these are corrected for the instrument response across the spectral window (bandpass correction), edited to remove radio frequency interference or unwanted atmospheric lines, and then averaged to form pseudo-continuum frequency bands. 

ALMA antennas measure two orthogonal senses of linear polarization simultaneously, $X$ and $Y$. All solar observations are currently made in dual-polarization mode. That is, the BLC produces correlations between antennas $i$ and $j$ as $X_iX_j$ and $Y_iY_j$ visibilities. The two linearly polarized correlations are generally summed to produce visibilities in total intensity (the Stokes I parameter). The two may be differenced as a means of estimating the thermal noise \citep{Shimojo2017}.The implementation of full polarimetry for solar observations is currently under study (\S5.1).

\subsection{Challenges Posed by the Sun}

The Sun poses a number of challenges for a general purpose instrument not specifically designed and optimized for solar observing. These challenges are over and above the already considerable challenges of observing in the mm/submm-$\lambda$ range. Among them are:

\begin{itemize}
\item The Sun is an intense emitter, so much so that it causes ALMA receivers to go into compression and for electronic elements along the signal path to saturate unless provisions are made. 
\item The Sun is a moving target, not only because of its apparent motion across the sky, but also because of its (differential) rotation.
\item The Sun is much larger than the FOV of a single ALMA antenna, which is $<20”$ at a wavelength of 1 mm, for example. 
\item Emission from the Sun is highly variable as a result of a variety of phenomena, ranging from chromospheric oscillations to flares.  
\end{itemize}
 
We now briefly summarize how each of these challenges have been addressed.

\subsubsection{Antenna Gain Reduction}

The ALMA antennas were carefully designed to allow them to safely point at the Sun. However, to avoid compression of the receivers and saturation along the electronic signal path, two steps are taken:

\begin{enumerate}

\item The gain of the SIS mixers in ALMA receivers is reduced by changing the voltage bias and/or the local oscillator (LO) current – the so-called “mixer detuned” or “mixer debiased” (MD) mode. The decrease in receiver gain provides greater headroom for the receivers to operate without going into compression. However, to ensure correct calibration transfer, both the Sun and calibrator sources must be observed in MD mode. 

\item There is still an enormous differential in the power entering the system when observing the Sun and observing calibrator sources, which are typically compact quasars or planets. Stepped attenuators along the signal path are used to set signal powers to optimum levels during an instrumental setup scan on the solar target. For all subsequent calibrator scans, the attenuators settings are reduced by a known and fixed amount on each antenna. In this way, robust phase transfer is ensured. 

\end{enumerate}

\noindent As shown by \citet{Shimojo2017} these modifications allow observations of quiet Sun and active region phenomena. Observations of solar flares are another matter, one that we discuss in \S5.4. A consequence of using the MD mode is an increase in system temperature which lowers the system sensitivity when observing calibrator sources, as least for bands 3, 5, and 6. For gain calibrators, it is highly desirable to observe a bright quasar ($>1$ Jy) relatively close to the Sun ($<15$ deg). Between late June and early July, there are no such bright quasars near the Sun, a period of time that should therefore be avoided for solar observing. In practice, it is rare that suitable array configurations are available for solar observing at this time because the larger configurations are favored during Chilean winter. 

\subsubsection{Apparent Source Motion}

The Sun and other solar system objects – planets, comets, asteroids – all display apparent motion. Their apparent position on the celestial sphere is described by an ephemeris. In the case of the Sun, the {\sl ALMA Solar Ephemeris Generator}\footnote{The ALMA Solar Ephemeris Generator was developed by Ivica Skoki\'{c} and is available at http:celestialscenes.com/alma/coords/CoordTool.html} is a convenient tool for preparing the solar ephemeris needed for observations on any specified date. The {\sl Ephemeris Generator} takes into account solar rotation via a standard or user-specified solar rotation model, as well as target offsets relative to the center of the solar disk. ALMA scheduling blocks link the user-specified solar ephemeris to the online system, allowing it to point and track the solar target of interest as discussed further in \S3.

\subsubsection{Mapping a Large Field of View}

The FOV of a single 12 m ALMA antenna is small compared to the angular diameter of the Sun, or even compared with the typical scale of an active region (few arcmin).  There are two issues:

\begin{itemize}
\item In order to increase the size of the angular domain mapped by ALMA, mosaicking techniques must be used \citep{Cornwell1989, Sault1996}). Mosaicking entails sequentially sampling a user-specified grid of array pointings to increase the effective FOV. The spacing between discrete pointings is set to ensure Nyquist sampling or better. Mosaic pointing patterns are referenced to the user-specified solar ephemeris. 
\item A related issue is that, as an interferometer, ALMA acts as a high-pass filter. While the maximum antenna baselines set the angular resolution of the array, the minimum antenna baselines determine the largest angular scales measured by the array. As seen in Table 3, the maximum recoverable scale measured by a given configuration of the 12 m array are all significantly less than the FOV! Two measures were implemented to mitigate this state of affairs:

\begin{itemize}
\item[i)] Solar observations use both the antennas in the 12-m array and those in the ACA as a combined array to perform interferometric (INT) observations. That is, all antenna baselines – 7 m x 7 m, 7 m x 12 m, and 12 m x 12 m are correlated in the BLC. In doing so, measurements on shorter baselines and larger angular scales are available and may be helpful in some circumstances (see \S4.1).

\item[ii)] Solar observers are provided with contemporaneous fast-scan total power (TP) maps of the full disk of the Sun \citep{Phillips2015, White2017}). These maps can be used to “fill in” the largest angular scales that are otherwise not measured by the interferometric array.
\end{itemize}
\end{itemize}

\noindent We return to some of the subtleties associated with these mitigation strategies in \S4.4. 

\subsubsection{Time Resolution}

Solar emission at mm-$\lambda$ and submm-$\lambda$ varies over a wide range of time scale, from $<1$~s to minutes, hours, and days. Depending on a given user’s science objectives, it may be necessary to resolve this time variability on the relevant time scale. If it is acceptable to average over time variability, one can exploit Earth rotation synthesis to improve sampling of the {\sl uv} plane somewhat with the available distribution of antennas. However, if the user wishes to resolve variable emission on time scales of minutes or less, the uv coverage does not change appreciably over the relevant time scale and the user is effectively constrained to snapshot {\sl uv} coverage. For many types of science, this trade-off is acceptable: the quality of the imaging is compromised but the source variability of interest is resolved. 

\section{ALMA Solar Observing Capabilities}

We now summarize ALMA’s solar observing modes and capabilities, current as of Cycle 9 in 2022-2023. We emphasize that development of additional modes and capabilities is ongoing, as discussed further in \S5. We note that while a number of software packages may be used to image ALMA data, the most powerful and well-supported is the {\sl Common Astronomical Software Applications}\footnote{https://casa.nrao.edu} (CASA) package. We make occasional reference to specific tasks or functions found in CASA. 

\subsection{Frequency Bands}

Four frequency bands are currently supported by ALMA for continuum solar observing (Table 1). Bands 3 and 6 were the first to be commissioned and were offered to the community in 2016 (Cycle 4). Testing and commissioning of solar continuum observations in ALMA Band 7 (346.6 GHz, 0.86 mm) were completed in 2018 and were made available to the community in 2019 as part of ALMA Cycle 7. Solar continuum observations in Band 5 (198 GHz, 1.51 mm) were commissioned in 2019 and were to be made available in 2020 for ALMA Cycle 8. Owing to the pandemic, however, Cycle 7 observations ended prematurely with the shutdown of the telescope in March 2020 and Cycle 8 was effectively canceled. The telescope was restarted in 2021 and the observing cycle number was reset to Cycle 8. Science observations recommenced in October 2021. 

Two MD modes were initially commissioned for Bands 3, 5, and 6. MD1 was initially designed for observations of the quiet Sun and MD2 for observations of active regions. In practice, however, it is found that MD2 results in more linear receiver performance although it comes at the cost of higher system temperature T$_{sys}$ ($\sim\!800$, 700, and 800 K, respectively). In the case of Band 7, no stable voltage de-bias settings were found. Hence, nominal receiver settings are used for band 7 observations although the stepped attenuators are optimized for solar observing. On one hand, T$_{sys}$ remains low ($\sim\!200$ K); on the other, the system is in mild compression ($\sim15\%$).

\begin{table}[ht]
\centering
\caption{ALMA Solar Observing Bands}
\medskip
\begin{tabular}{c | c | c | c | c | c | c }
\hline
Freq. Band & $\nu_{LO}$ (GHz) & $\lambda_{LO}$ (mm) & spw~0 (GHz) & spw~1 (GHz) & spw~2 (GHz) & spw~3 (GHz) \\
\hline
3 & 100 & 3 & 92-94 & 94-96 & 104-106 & 106-108 \\
5 & 198 & 1.51 & 190-192 & 192-194 & 202-204 & 204-206 \\
6 & 239 & 1.25 & 229-231 & 231-233 & 245-247 & 247-249 \\
7 & 346.6 & 0.86 & 338.6-340.6 & 340.6-342.6 & 350.6-352.6 & 352.6-345.6 \\
\hline
\end{tabular}
\end{table}

\subsection{Array Configurations}

Four of the ten possible 12-m array configurations are available for solar observing as shown in Table~2. Check marks indicate that the configuration is available for solar observations.  There are two reasons for this restriction: first, at mm-$\lambda$ and submm-$\lambda$ precipitable water vapor in the sky introduces phase variations to the wavefront incident on each antenna that can propagate into the visibilities as image motion and/or a loss of coherence unless corrected. ALMA uses water vapor radiometers (WVRs) on each antenna to measure brightness temperature variations of the sky, a proxy for the phase variations \citep{Nikolic2013} that may be used to make antenna-based phase corrections. Unfortunately, the WVRs saturate when pointing at the Sun and such corrections are therefore not available for solar observations. Instead, self-calibration techniques must be used to correct phase variations (\S5). However, for self-calibration to be effective, the loss of coherence due to phase fluctuations, which increases with baseline length, must not be so extreme as to render the technique ineffective because it relies on a plausible initial model of the source. Hence, the maximum array configurations in each available frequency band is chosen to ensure that self-calibration techniques can be used in most cases to correct for phase fluctuations introduced by the sky. 

\begin{table}
\begin{center}
\center{\caption{ALMA Array Configurations for Solar Observing}}
\medskip
\begin{tabular}{l | l | c | c | c | c }
\hline
& Band & 3 & 5 & 6 & 7 \\
\hline
Configuration & $\nu$ (GHz)/$\lambda$ (mm) & 100/3 & 198/1.51 & 239/1.25 & 346.6/0.86 \\
\hline
&&&&&\\
ACA & b$_{min}$ (m): 8.7 & \checkmark & \checkmark & \checkmark & \checkmark \\
 & b$_{max}$ (m): 45.0 & & & & \\
 \hline
 &&&&&\\
C-1 & b$_{min}$ (m): 14.6 & \checkmark & \checkmark & \checkmark & \checkmark \\
 & b$_{max}$ (m): 160.7 & & & & \\
 \hline
 &&&&&\\
C-2 & b$_{min}$ (m): 14.6 &  \checkmark & \checkmark & \checkmark & \checkmark \\
 & b$_{max}$ (m): 313.7 & & & & \\
 \hline
 &&&&&\\
C-3 & b$_{min}$ (m): 14.6 &  \checkmark & \checkmark & \checkmark &   \\
 & b$_{max}$ (m): 500.2 & & & & \\
 \hline
 &&&&&\\
C-4 & b$_{min}$ (m): 14.6 &  \checkmark &  &  &  \\
 & b$_{max}$ (m): 783.5 & & & & \\
\hline
\end{tabular}
\end{center}
\end{table}

A second issue is that as the array configuration increases in size with a fixed number of antennas, sampling in the {\sl uv} plane becomes increasingly sparse. Since {\sl uv} coordinates are measured in wavelength units, it is also the case that the sampling becomes increasingly diluted with increasing frequency.  This is readily seen in Table~3 which shows the array resolution and the maximum recoverable scale\footnote{The maximum recoverable scale is $\theta_{MRS}\approx 203\lambda_{mm}/L_5$ arcsec, where $L_5$ is the radius in meters within which 5\% of the baselines in a given configuration reside. $L_5$ is 9.1, 21.4, 27.0, 36.6, and 54.1 for the ACA, C-1, C-2, C-3, and C-4 configurations, respectively.}  $\theta_{MRS}$ for each frequency band and configuration. While the angular resolution improves for a given configuration as one moves to higher frequencies, or as one increases the size of the array configuration at a fixed frequency, the price paid is poorer sampling of the Fourier domain. This is an important consideration when imaging a target like the Sun, which emits on angular scales ranging from the available resolution to scales much larger than the FOV of a given antenna. The configurations offered for solar use therefore represent a compromise between angular resolution and the density of {\sl uv} sampling. 

\begin{table}
\begin{center}
\center{\caption{ALMA Angular Resolution and Maximum Recoverable Scale}}
\medskip
\begin{tabular}{l | l | c | c | c | c }
\hline
& Band & 3 & 5 & 6 & 7 \\
\hline
Configuration & $\nu$ (GHz)/$\lambda$ (mm) & 100/3 & 198/1.51 & 239/1.25 & 346.6/0.86 \\
\hline
ACA &$\theta_{res}$(arcsec) & 11.6 & 6.77 & 5.45 & 3.63 \\
 & $\theta_{MRS}$ (arcsec)  & 66.0 & 36.1 & 29.0 & 19.3 \\
 \hline
C-1 & $\theta_{res}$(arcsec) &  3.38 & 1.83 & 1.47 & 0.98 \\
 &  $\theta_{MRS}$ (arcsec)  & 28.5 & 15.4 & 12.4 & 8.25 \\
 \hline
C-2 & $\theta_{res}$(arcsec) &  2.30 & 1.24 & 1.00 & 0.67 \\
 & $\theta_{MRS}$ (arcsec)  & 22.6 & 12.2 & 9.81 & 6.54 \\
 \hline
C-3 & $\theta_{res}$(arcsec) &  1.42 & 0.77 & 0.62 &  \\
 & $\theta_{MRS}$ (arcsec)  & 16.2 & 8.73 & 7.02 & \\
 \hline
C-4 & $\theta_{res}$(arcsec) &  0.92 &  &  &  \\
 &$\theta_{MRS}$ (arcsec)  & 11.2 & & & \\
\hline
\end{tabular}
\end{center}
\end{table}

It is important to note, however, that any of the four 12-m array configurations available to solar observers may be used in combination with the ACA 7-m array. That is, solar observations make use of a single heterogeneous array that includes available 12-m array antennas and ACA 7~m antennas. The reason for doing so is readily apparent from Table 3: $\theta_{MRS}$ for any band and any configuration is less than the FOV. Inclusion of the ACA 7 m antennas provides sampling up to angular scales comparable with the nominal FOV; and as previously noted, full-disk total power maps are also available to fill in emission on the largest angular scales. We return to issues related to the use of the heterogenous array in \S4.1. 

\subsection{Time Resolution}

The time resolution initially offered to the solar community for Cycle 4 was 2~s. This was reduced to 1~s beginning with Cycle 7. ALMA provides a much higher snapshot imaging cadence than is typically available at UV/EUV wavelengths from space missions such as the {\sl Solar Dynamics Observatory}(SDO; 12 s) and the {\sl Interface Region Imaging Spectrograph} (IRIS; 20 s). A study led funded by the European Southern Observatory is exploring opportunities for even higher time resolution imaging, as well as the use of larger antenna configurations to enable higher angular resolution imaging (\S5). 

\subsection{Mosaicking}

The mosaicking technique was introduced in \S1.2.3. The upper limit to the number of pointings in a given mosaic is currently limited to 150 for both solar and non-solar observations. An example of a large format mosaic is shown in Fig.~2. The minimum time per pointing for solar observations is 7.6 s, including overhead for moving the antennas from one point to another. Therefore, including calibration, more than 20 min is required for a single large format mosaic. More limited mosaics can, of course, be executed but users must weigh the tradeoffs between the desired FOV, the time required to execute a given mosaic, and the time scale on which the source may change. An example of a ten-pointing mosaic is shown in \S4.1.

\begin{figure}
\begin{center}
\includegraphics[angle=0,trim={0cm 4cm 0cm 0cm},clip,width=8cm]{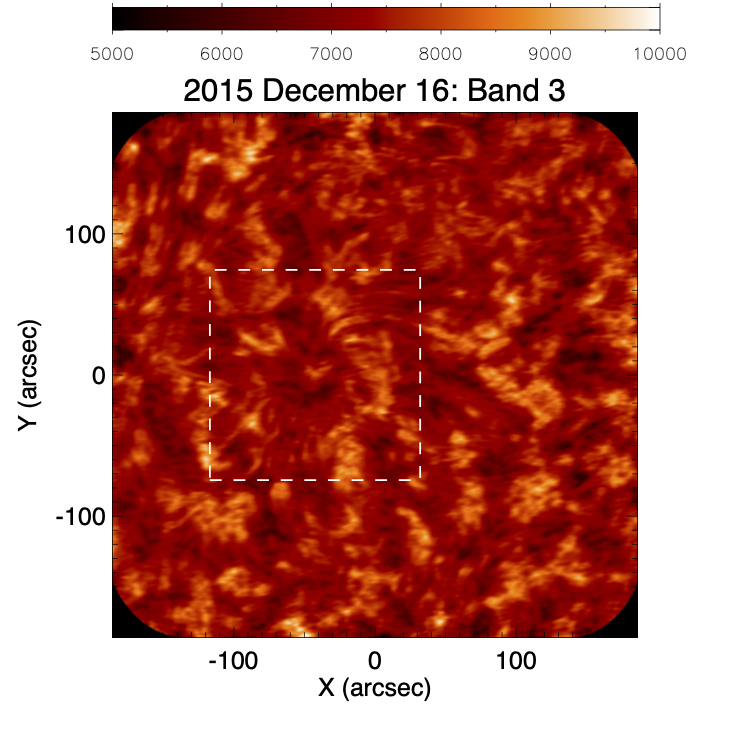}
\includegraphics[angle=0,trim={0cm 4cm 0cm 0cm},clip,width=8cm]{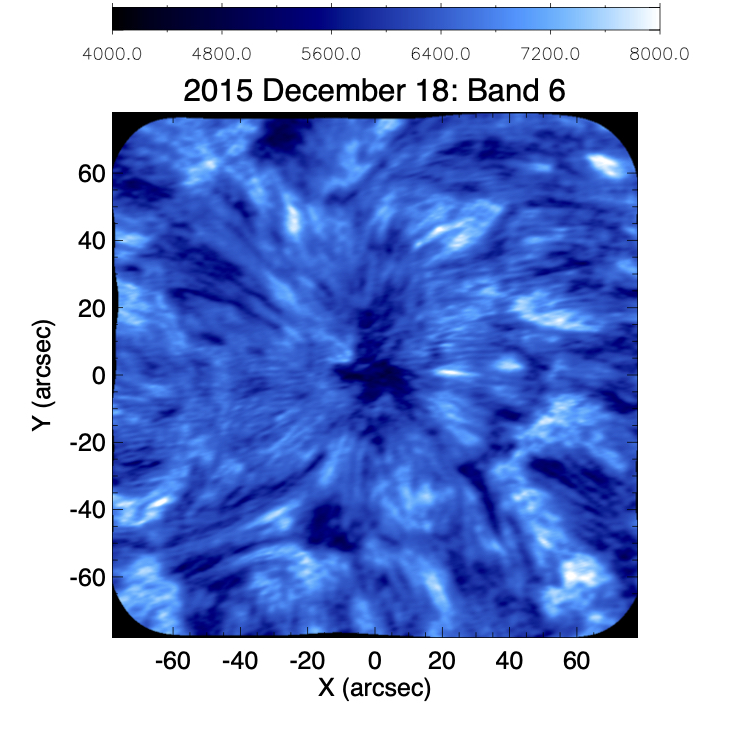}
\end{center}
\caption{Examples of large-format mosaics from ALMA science verification data. Left panel: a 149-pt mosaic in Band 3 (3~mm) of a sunspot-containing active region. Right panel: a 149-pt mosaic in Band 6 (1.25~mm) of the same sunspot two days later. The equivalent field of view of the Band 6 image is shown on the Band 3 as a dashed-line box. Note the difference in brightness temperature scale. See also \citet{Shimojo2017}.}\label{fig:2}
\end{figure}

\subsection{Fast-scan Total Power Mapping}

Full-disk TP maps supplement interferometric (INT) observations. These provide broader context images for the INT data and provide the means of ``filling in" short {\sl uv} spacings, critical if absolute brightness temperatures are needed. The most straightforward technique for combining TP and INT data is ``feathering" \citep{Cotton2015}. See \S4.4 for further discussion about combining TP and INT data. 

As described by \citet{White2017} a double-circle pattern is employed to scan a region centered on the Sun that is 2400" in diameter (see Fig.~3 for an example). The resolution of the map is determined by the resolution of the 12~m total power antenna: 58.3", 29.3", 24.3", and 16.7" at Bands 3, 5, 6, and 7, respectively. The time required to produce a full disk map depends on the frequency band: e.g., 13~min for Band 3 and 17~min for Band 6. Calibration of the full disk maps is challenging. Non-repeatability between specific instances led \citet{White2017} to suggest best-values for the quiet Sun brightness temperature at the center of the solar disk based on an aggregation of many test observations. Recently, \citet{Alissandrakis2022} performed an analysis of the center-to limb brightening of ALMA full-disk TP maps in Bands 3, 6, and 7 and showed that Bands 6 and 7 could be cross-calibrated against Band 3 in a self-consistent manner. Table~4 shows the quiet Sun brightness temperature originally recommended by \citet{White2017} and those resulting from the analysis of \citet{Alissandrakis2022}. Ultimately, quiet Sun brightness measurements should be referenced to an independent standard such as the Moon. 

\begin{table}
\begin{center}
\caption{Total Power Map Scaling}
\medskip
\begin{tabular}{l | c | c }
\hline
Band & White et al. (2017) & Alissandrakis (2020) \\
\hline
3 & 7300 K& 7347 K\\
5 & & 6532 K\\
6 & 5900 K& 6347 K\\
7 & & 6085 K\\
\hline
\end{tabular}
\end{center}
\end{table}

The time required to produce full-Sun maps (10-20~min) is typically longer than the time scale of many dynamic phenomena of interest. There is also a significant mismatch between the time resolution available for INT data (1~s) and the TP mapping time.  It is not possible to drive the TP antennas at higher rates to produce higher-cadence maps and so the field of view must be reduced to decrease the mapping time. In Cycle 9, therefore, a second fast-scanning TP mapping mode has been made available to solar observers: Fast Regional Mapping (FRM). The intent of the FRM mode is the same as that of the full-Sun maps: to obtain short-spacing data that is complementary to the INT data, but on a time scale that is more closely commensurate with changes in the source. Therefore, the coordinates of the field center are the same as the target coordinates of the interferometric observation. Moreover, the scan pattern of the FRM released in ALMA-Cycle 9 is the double-circle pattern, which is the same as for the full-Sun map, and the shape of the field of view is a circle. The observer only needs to specify the need for FRM observations in support of a particular science goal, and specify the diameter of the mapping domain. Additional scan patterns are under development: the use of a Lissajous scan patterns allows rectangular mapping domains.  

\begin{figure}[h!]
\begin{center}
\includegraphics[trim={0cm 1.5cm 0cm 2cm},clip,width=16cm]{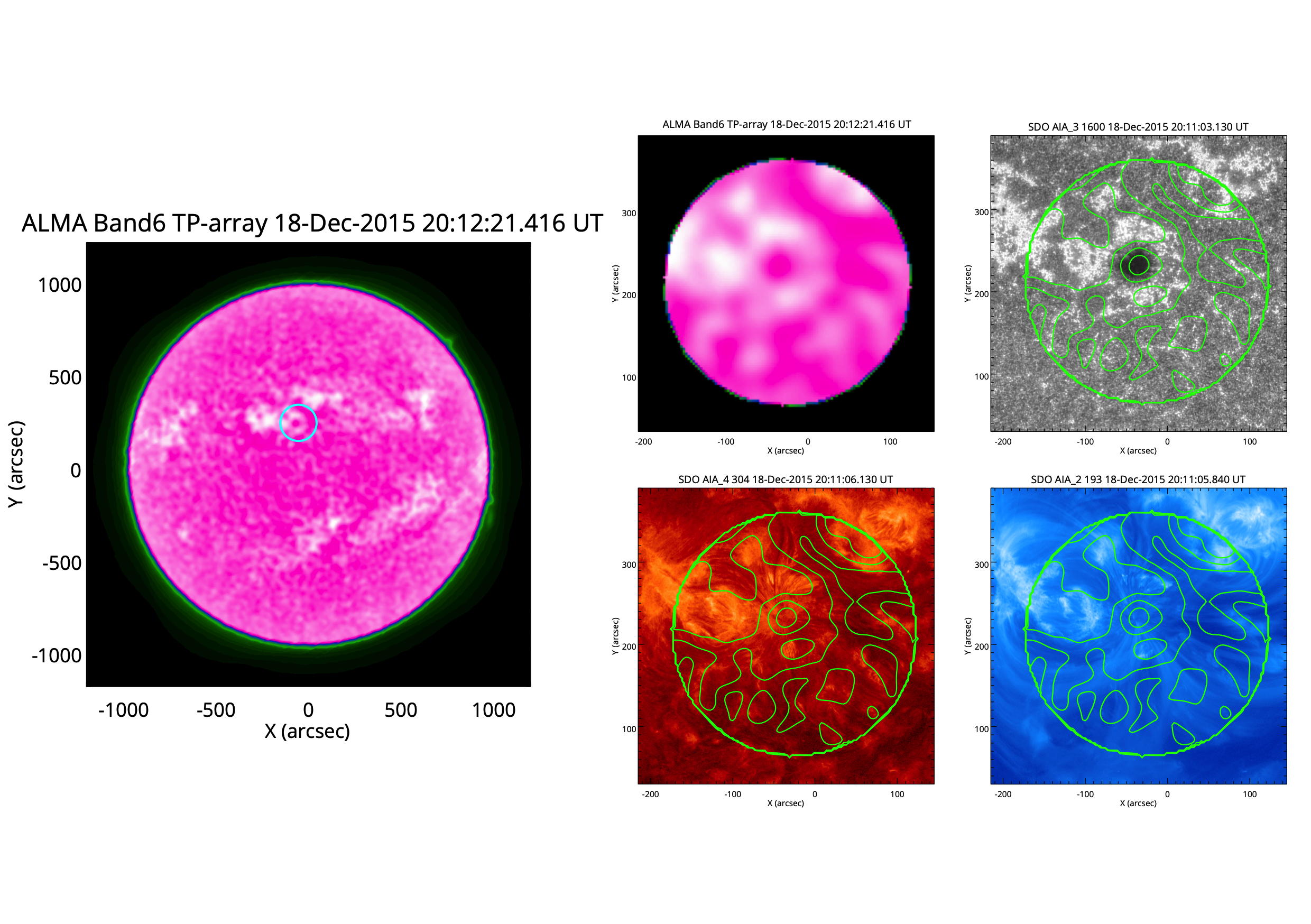}
\end{center}
\caption{Simulated FRM data. Left panel: Full disk TP image obtained with Band~6 on 18 December 2015. The circle indicates the field of view of FRM with 300” diameter. Right panels: Simulated FRM image and UV/EUV images. Upper Left: Simulated FRM image with 300” diameter FoV. Upper Right: UV continuum (1700Å band). Lower Left: He II 304Å Band, Lower Right: Fe XII 193 Å Band. Green contours show the FRM image.Examples of a total power full disk map (left) and FRM map (middle, top). The Band 6 total power map was acquired during commissioning in December 2015. The FRM map was made using the TP map in simulation}\label{fig:3}
\end{figure}

The size of the field of view and the mapping cadence is a trade-off that depends on the science objectives. Test observations in Band 6 found that map durations are 11~s, 21~s, 32~s, and 63~s for map diameters of $100", 200", 300"$, and $600"$, respectively, consistent with detailed simulations. FRM maps can therefore be obtained on a cadence comparable to SDO or IRIS imaging observations, for example. FRM observations are available simultaneously with INT observations but, similar to INT observations, FRM observations must be periodically interrupted for calibration.  In addition, at beginning and end of an FRM observation, full-Sun maps are obtained for flux calibration. Therefore, an FRM observation always includes numerous regional maps and two full-disk maps. The FRM maps are calibrated against the full-disk maps in two steps: the full disk brightness is scaled to the recommended quiet-Sun value near the center of the the disk and the FRM maps are then scaled from the full-disk maps at the appropriate offset.  Since it is assumed that the brightness of the reference point does not change significantly during the observations, the diameter of the FRM mapping domain should be chosen to include quiet Sun as the reference point. Hence, a mapping region $\ge 200"$ is recommended, in general. 

\subsection{System Issues}

Unfortunately, a number of system issues have affected solar data at various times to varying degrees. All issues have related to pointing or tracking. Once identified, most issues have been addressed and subsequent observations have been unaffected. From a scientific perspective, it has usually been possible to recover from the errors introduced by these issues but it is important for users, especially users of archival data, to be aware of them. The first affects Cycle 4 data observed in December 2016. Instead of tracking the user-specified ephemeris the system simply tracked the center of the solar disk; i.e., it did not take any user-specified target offsets into account or track with solar rotation. At this time, the Sun was in solar minimum and most programs were to observe the quiet solar chromosphere. Some observers were able to correct their data for the lack of rotational tracking and address their science objectives but for others their science objectives were compromised. The problem was corrected and additional operational checks were put into place to verify that targeting is indeed as requested by the PI. 

\begin{figure}[h!]
\begin{center}
\includegraphics[trim={0cm 4cm 0cm 4cm},clip,width=15cm]{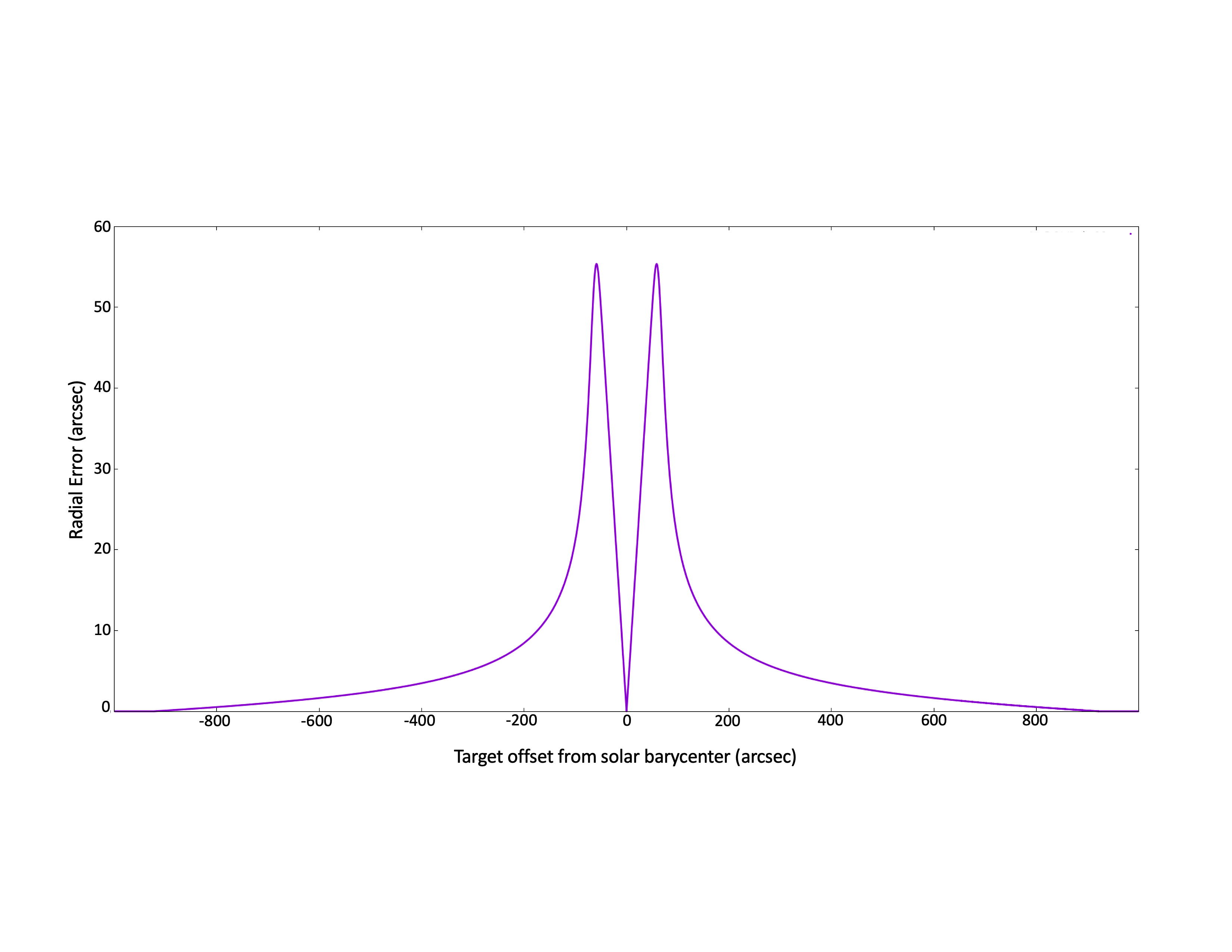}
\end{center}
\caption{The radial position error of the target as a function of target offset from the solar barycenter.}\label{fig:4}
\end{figure}

A second issue is related to the reference coordinates assigned to a given observations. As an ephemeris object, the geocentric coordinates of the source change continuously in time. Hence, a reference coordinate corresponding to the appropriate reference time is assigned. For observations in Cycles 4 and 5, the coordinates were determined using the CASA task {\sl fixplanets} using the antenna POINTING table in measurement set. These were found to be inaccurate by as much as 30". Since Cycle 6 the ephemeris table is used directly using the CASA {\sl phasecenter} function of the {\sl msmetadata} tool ({\sl msmd.phasecenter()}. Archival data should be processed using up-to-date scripts.
%A simple case study has been performed which compared archive images of Cy 4 with their counterparts - the same data just re-processed by the upgraded calibration script. The comparison shows a displacement up to 30 acrsec between old and new method for attributing the celestial coordinates to the science-target field. On the other hand, no significant displacement is seen between the brightness structures in the newly reprocessed solar INT images and their, e.g., optical counterparts.

In Cycle 7, residual pointing effects were traced to general relativistic (GR) corrections for light bending near the Sun\footnote{See the knowledgebase article at https://help.almascience.org/kb/articles/what-is-a-cycle-7-solar-data-issue-i-should-be-aware-of for details.}. The most important of these occurred as the result of the misapplication of GR delay corrections for target offsets within 920" of the solar barycenter. The error results in a radial pointing offset that is a function of the desired target offset. Fig.~4 shows that the radial pointing offset increasing linearly with target offset from 0" to $\approx\!55"$ and then decreases with increasing target offset, dropping below 5" for radial offsets $>5'$. The error is not present for targets on the solar limb. This problem persisted until March 2020 but has now been corrected. Users should use source names beginning with ``Sun" (case insensitive) to ensure that GR corrections are disabled for solar observations.  In many cases, the offset introduced by this issue does not seriously compromise science objectives because the pointing error is $<10"$ in many cases. However, if ALMA images are being compared with data from other missions and observatories they must obviously be corrected in order to properly co-align the data.  For mosaics, it appears that while the reference pointing established by the ephemeris is affected, offset pointings relative to the reference were not. 

A related error originates in the 30~s cadence at which GR delay corrections were made. For an ephemeris object like the Sun there is a drift of the target relative to the pointing offset that is ``reset" with every GR delay event, resulting in a pointing ramp every 30~s. The magnitude of the effect is of order $1"$ and can be corrected through self-calibration or cross-correlation. Fig.~5 shows an example. Not only were interferometric data affected by the GR delay corrections, antenna pointing was, too. The maximum antenna pointing error occurs at the limb and is 1.85". It has no impact on source positions and is small compared to the primary beam in Band 3 but can affect the primary beam correction for Bands 6 and 7.

\begin{figure}
\begin{center}
\includegraphics[trim={1cm 10cm 1cm 1cm},clip,width=12cm]{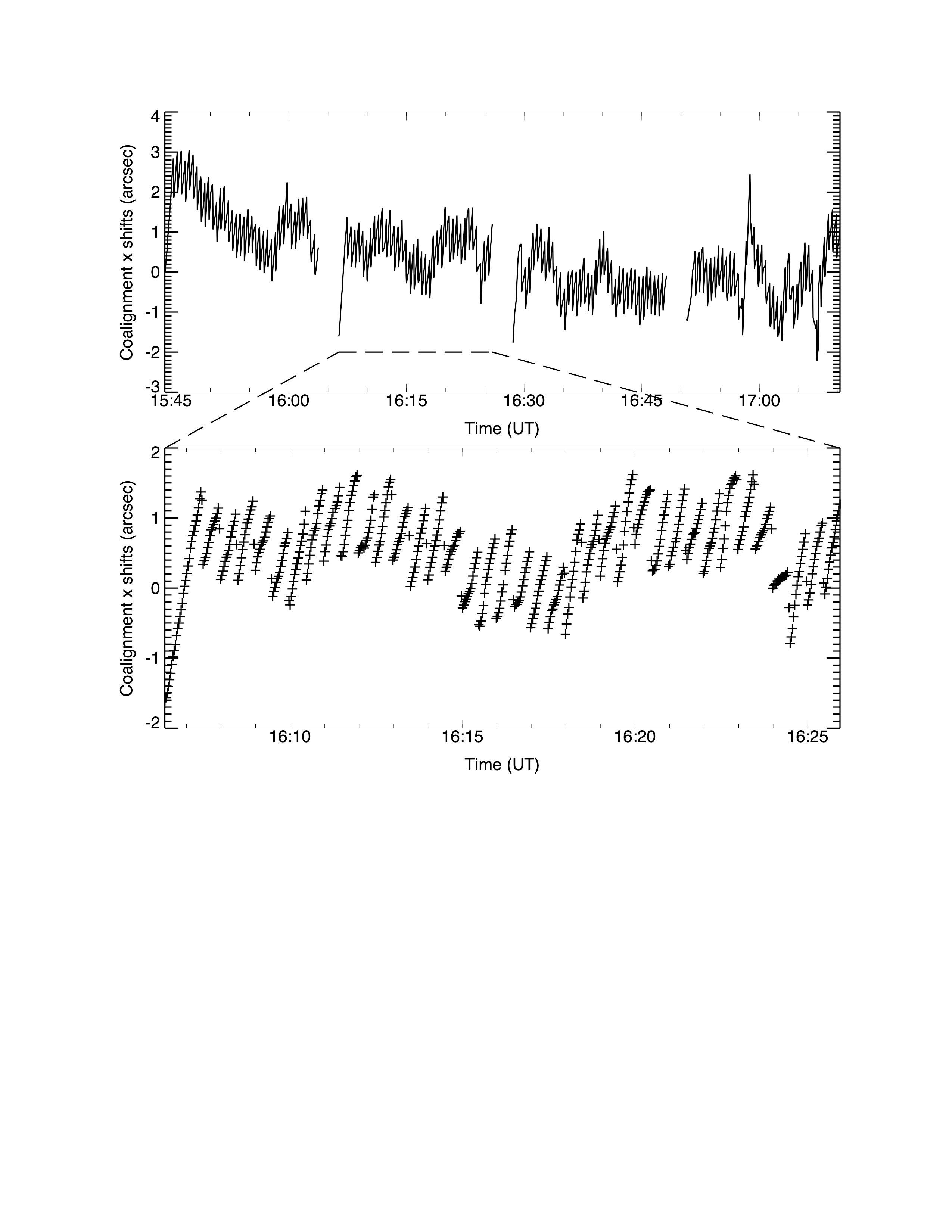}
\end{center}
\caption{Example of the pointing ramping that results from the application of the GR delay tracking correction every 30~s. The top panel shows the change in pointing for the duration of an SB. The bottom panel shows a detail in which the ramping is vividly illustrated. }\label{fig:5}
\end{figure}

Finally, full disk TP maps have been subject to small timing errors that manifest as low-level artifacts at the solar limb as pointed out by \citet{Alissandrakis2017}. This will soon be corrected with the use of higher-performance 12~m antennas for TP mapping and updates to the relevant software. 

\section{ALMA Solar Program Workflow}

The life-cycle of any ALMA project, from an initial idea up to publication of the results, is schematically drawn in Fig.~3. Here, we discuss elements of proposal preparation, proposal review, solar observing, data calibration, and quality assurance (QA). Issues related to data imaging and analysis are discussed in \S4.

\begin{figure}[h!]
\begin{center}
\includegraphics[angle=0,width=15cm]{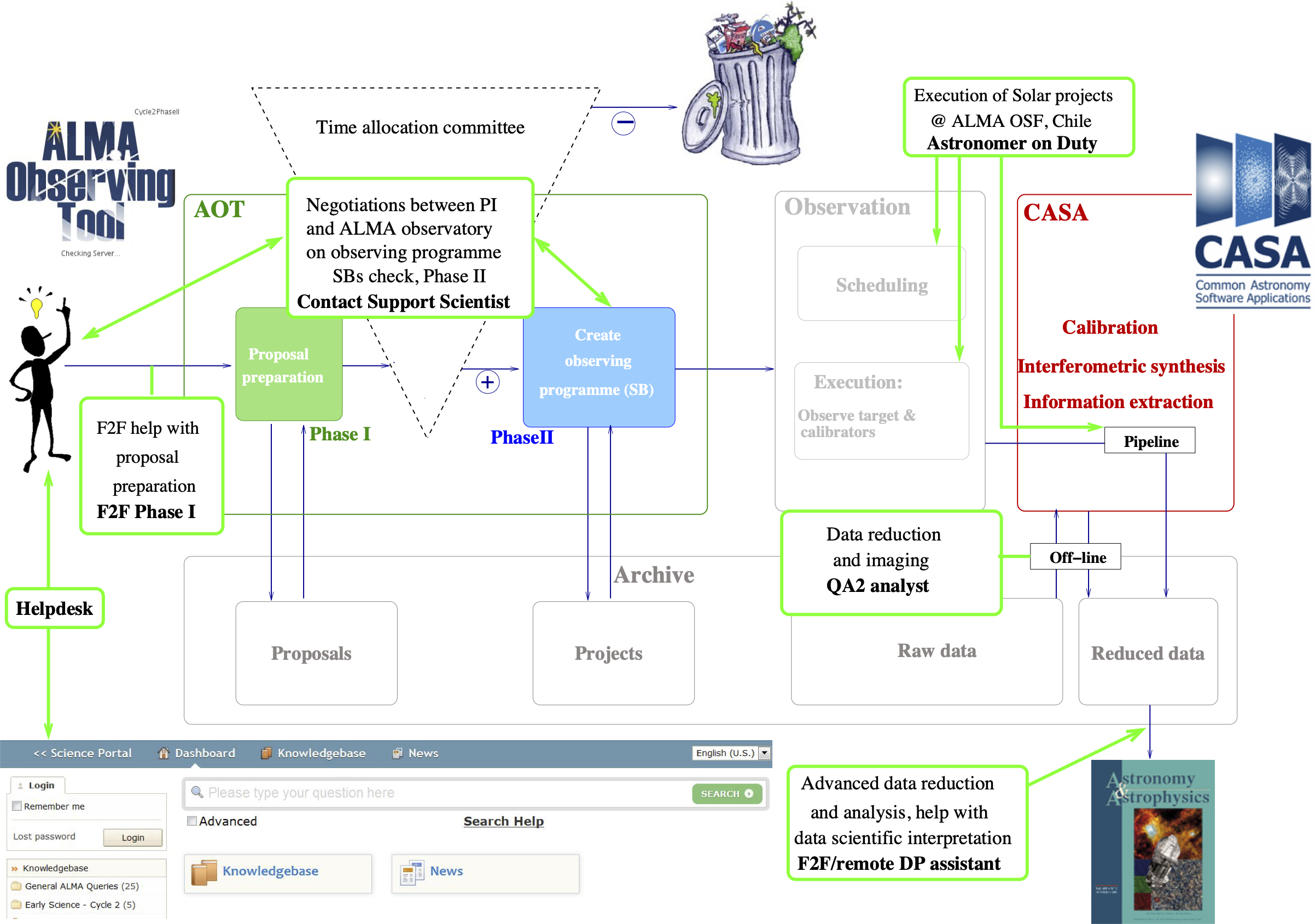}
\end{center}
\caption{Schematic representation of the workflow of a project from proposal development and submission to data reduction and analysis.}\label{fig:6}
\end{figure}

The life-cycle of a given science program formally begins by with the submission of a proposal for observing time in response to an ALMA call for proposals. Calls are issued on an annual basis in March and the proposal deadline is in April. Proposal preparation and submission is referred to as Phase I. Investigators are notified of the disposition of their proposal in August. Should the proposal be successful in the subsequent evaluation (see \S3.3 below), so called Scheduling Blocks (SBs) are generated from technical details specified in the proposal. This procedure works mostly automatically, with a small adjustments made by the Contact Scientist (CS) and the observatory staff in cooperation with the project Principal Investigator (PI). This is called Phase II. An SB is later converted to a Python script loaded by ALMA during observations. An observing cycle lasts one year, beginning on October 1. Solar observations are only possible with a limited number of array configurations (Table 3) and so users should be aware of the array configuration schedule\footnote{ALMA antenna configuration schedules can be found at https://almascience.eso.org/observing/observing-configuration-schedule.} during the course of the year. 

The primary interface between ALMA and the user is an ARC. Depending on nationality, users will be affiliated with a given ARC. From the user point of view there are three important components of the ARC support: the Helpdesk system, the Contact Scientist (CS), and Data Analysts (DAs). While the Helpdesk is open to user queries anytime it also contains a rich knowledgebase that can be consulted.  The CS is an ARC astronomer assigned to a given project when it is accepted, i.e. before the Phase II starts. The communication between the PI and CS starts by opening a project-dedicated Helpdesk ticket – the ARC will do so on PI’s behalf. Following the successful execution of an observation, the relevant ARC is responsible for calibrating the data and performing quality assurance. DAs perform these functions as we discuss further below. 

%When the solar observations of a new cycle come, it depends largely on the array configuration schedule – recall that we are limited to the most compact array geometries (Table 3). 

%Generally speaking, given configuration limitations, the periods suitable for solar observations come in December/early January and then in late March/April or even May. In those appropriate periods, concentrated into 3-4 weeks lasting solar campaigns, most of the day time is reserved for the solar observations. The PIs are notified 1 month ahead the campaign starts in order to participate on formation of the campaign observing schedule

Solar observations with ALMA often coordinate with other ground-based or space-based telescopes and observatories. Hence, careful coordination between the PI and JAO science operations is necessary (\S3.4) After the observations are executed by JAO science operations the raw data are transferred to the ARCs for processing. This includes calibration, reference imaging, and quality assurance (QA) assessment. This process typically lasts several weeks following an observations. Should the data pass the QA, they are delivered  to the PI. The PI receives (1) raw data, (2) calibration and imaging scripts (Python) together with the master scriptForPI.py, and the products – FITS files with the images, one per each Execution Block, both for interferometric (INT) and total power (TP) observations. Additional post-processing (e.g. INT + TP data combination, time-domain imaging with self-calibration loop, etc.) is the PI’s responsibility; however, the ARCs can provide assistance and in some cases and the advice of ARC astronomers can be solicited at any time via the ALMA Helpdesk system. 

\subsection{Proposal Preparation and Submission}

An ALMA proposal is prepared and submitted using the ALMA Observing Tool (OT)\footnote{The ALMA OT, as well as the user manual and online tutorials, is available at https://almascience.nrao.edu/proposing/observing-tool or at https://almascience.eso.org/proposing/observing-tool}. It is a Java-based GUI application. In brief, it contains a set of query pages where the PI describes the project setup: 

\bigskip
\begin{itemize}
\item Basic information about the project: PI, co-investigators, title, scientific category. There are five science categories:
\begin{enumerate}
\setlength{\itemindent}{0.25in}
\item Cosmology and the high redshift universe 
\item Galaxies and galactic nuclei 
\item ISM, star formation and astrochemistry 
\item Circumstellar disks, exoplanets and the solar system 
\item Stellar evolution and the Sun 
\end{enumerate}
\item Science goals (SGs): one per target and observing band (e.g. “Prominence in Band 3”). For each SG one must provide:
\begin{itemize}
\setlength{\itemindent}{0.25in}
\item[--] Spectral setup 
\item[--] Field setup 
\item[--] Control and performance 
\item[--] Technical justification. 
\end{itemize}
\item Scientific justification
\end{itemize}
\medskip

\noindent Proposers are encouraged to consult the ALMA Proposer's Guide\footnote{The ALMA Proposer's Guide is updated each observing cycle and is available at https://almascience.eso.org/documents-and-tools}. As noted in \S2.6, users should prefix each source name with ``Sun" to ensure that the system handles relativistic corrections correctly. We briefly summarize key elements of an SG. 
\smallskip

\noindent The {\sl Spectral setup} currently allows only the  “single continuum” frequency bands specified in Table 1. The spectral windows within a given frequency band are fixed; i.e., one cannot change the LO frequency to place them elsewhere in the frequency band. Observations of the same target may be performed in more than a single band – for each, a separate SG must be defined. ALMA cannot observe multiple frequency bands simultaneously (see \S5). The best that can be done at present is to perform back-to-back observation of the same field consecutively in different frequency bands. 
\smallskip

\noindent The {\sl Field setup} for each SG can be more complicated: one can chose between single pointing (SP) and a mosaic. An SP has a FOV approximately defined by the size of the primary beam of a 12-m antenna. Should a larger FOV be required to meet science objectives, a mosaic must be requested  (\S\S1.2.3, 2.4) and the user must specify the angular size of the domain to be imaged. The OT automatically calculates the number of discrete antenna pointings required in the mosaic. Nyquist sampling is the default spacing between pointings.  Any unique combination of spectral and field setup requires a separate SG. Beginning in Cycle 9 (2022) the fast regional mapping (FRM; \S2.5) capability became available. The OT allows users to define the circular area to be scanned with relatively high cadence encompassing the interferometric image. To avoid the Fourier aliasing, it is recommended that users select a region that is twice the size of the interferometric FOV. For the purposes of proposal preparation, a representative solar ephemeris must be uploaded for each SG in the field setup. As described below, the ephemeris actually used by a given solar observing program must be provided one or two days before a given program is executed.  All target pointings -- SP, mosaic, and FRM -- are referenced to the user-provided ephemeris.  
\smallskip

\noindent {\sl Control and performance} parameters in the OT are used by the user to specify the desired angular resolution or an acceptable range of angular resolutions for the project.  Coupled with the desired spectral setup, the appropriate array configuration(s) will be determined.  
Sensitivity parameters for an SG can often be ignored since there is no lack of signal on solar targets, in general. Exceptions do occur, however; for example, solar prominences at high frequencies, as they can be quite optically thin. The default integration time of 2~s is sufficient in most cases. The maximum time of a single realization of the observation of the given SG (the so-called Execution Block – EB; see below) is 2 hours. Note, that this period means $\sim\!1.5$ hour on the science target, the rest is spent on instrument setup and calibration.
\smallskip

The {\sl Technical justification} is used to justify each SG setup and to demonstrate its feasibility. It is highly recommended that proposers specify and justify any requests that may have an operational impact: for example, that back-to-back execution of particular SBs is required, or that coordination with another telescope will be needed. An excellent resource for technical aspects of ALMA is the Technical Handbook\footnote{The ALMA Technical Handbook is available at https://almascience.eso.org/documents-and-tools}, which is updated each observing cycle. 
\medskip

The key component of an ALMA proposal is the scientific justification. It is of critical importance as it makes the scientific case for an allocation of telescope time to address the stated science goals. Solar proposers should be aware that not all reviewers are necessarily solar experts. The proposal should provide enough context and background for a non-expert to understand the science goals. The scientific justification is the largest determinant in whether a given proposal receives a time allocation!

Before a proposal can be submitted it must be ``validated". That is, it is checked by the OT for completeness and feasibility within the constraints imposed on solar observing programs. For example, the validator does not allow you to observe the Sun in Band 9 or to use C-7 array configuration (as implied from selecting an angular resolution that requires an array configuration that is not allowed). Upon validation, a proposal may be submitted. 

\subsection{Proposal Review}

Before 2021 a large number of ALMA Review Panels (ARPs) was allocated to the five different science categories based on proposal pressure within each category. Typically a minimum of 2 ARPs and a maximum of 6 was allocated to each category to review and rank each proposal within that category. The ALMA Program Review Committee (APRC), comprising APR Chairs, then convened to perform a final review of proposal rankings and to determine which large proposals would be accepted. Large proposals are those requiring allocations of more than 50 hrs of time on the 12-m array.  

Beginning with Cycle 8 (2021) the JAO adopted a Distributed Peer Review (DPR) system. The process is ``double blind": reviewers do not know the identities of the proposing team and proposers do not know the identities of proposal reviewers. All proposals are evaluated and ranked using the DPR process with the exception of large proposals, which continue to be evaluated by the APRC. All PIs, or a designated co-investigator, must review ten proposals for each proposal they submit\footnote{Additional details regarding DPR process and policies may be found at https://almascience.nrao.edu/documents-and-tools/cycle9/principles-review-process}.  Since a given solar proposal is being evaluated by peers who also submitted proposals, not all of whom are solar experts, it is important to ensure that a proposal is understandable to non-experts. 

Most proposers are affiliated with one of the four ALMA partners. Those who are not may still propose for observing time under ``open skies". A total of 5\% of the available observing time is available to unaffiliated observers\footnote{See https://almascience.nrao.edu/documents-and-tools/cycle9/alma-user-policies for additional details regarding user policies}. ALMA typically receives 1700-1800 proposals for each observing cycle. Of these, only a few dozen proposals are typically submitted by the solar community. The instrument is highly over-subscribed with only 15-20\% of submitted proposals approved for a time allocation and so the number of solar programs observed to date is not large. 

\subsection{Solar Observing Process}

Solar observing in practice has evolved since first becoming available in late-2016. Initially, solar programs were executed in a ``campaign mode". While this was a convenient and flexible way to maximize the number of programs scheduled within the campaign period, they relied on the presence of solar scientists at the ALMA OSF. A reliance on external solar scientists, especially if several array configurations were required, was deemed unsustainable.  In more recent years, the execution of solar programs is coordinated through the PI, the CS, and the Astronomer on Duty (AoD). 

%Experience from the already gone observing cycles shows that the best success rate for solar ALMA observations is reached, when they are performed in the Solar Campaign mode. This means that from the available time windows based on suitable array configurations (Table 1), a few weeks (usually three or four) are allocated for the continuous solar observations during the day time. The solar campaign is interrupted only by the engineering time (predictable and accounted in the campaign observing schedule) and high-priority “Target of Opportunity” projects, including (usually partly overlapping) VLBI campaign. The exact amount of time allocated for the campaign depends on the number of solar projects accepted. As a rule of thumb, the campaign is split into two parts – Winter and Spring. Once the time window for the solar campaign is allocated and approved by JAO (known to the ARCs usually in October-November), there is time to divide the allocated slots for the specific projects – create the Campaign observing schedule. This happens basically by negotiation between the ARC solar-team representatives (EU, NA, and EA – see above) and the JAO.

Several days before a given observation can potentially be executed, the PI opens a ``Target of Opportunity” (ToO) ticket using the Helpdesk, in which the PI briefly describes the details of observation, including the target coordinates, whether to track solar rotation, and any constraints that might impact operations – e.g., whether possible loss of complementary TP maps compromises the science goal or is tolerable, whether  back-to-back observations of some blocks are absolutely essential or not, etc. The ToO ticket should be updated before each potential execution of an SB. The AoD is currently responsible for generating the solar ephemeris based on information provided by the PI in the ToO ticket, and for any changes to the orientation of a mosaic, if one or more is requested. Before the actual observation takes place, its execution is simulated, the results of which are reviewed and approved by the PI and CS. If a given SG is being supported by an external observatory or mission, the PI is responsible for coordinating such support. 

As described above, each SG defined in the OT proposal (Phase I) is automatically transformed into an SB during the Phase II. The SBs represent a complete observation setup but critical information must still be provided. As noted, an ephemeris must be generated shortly before an SB is executed. In addition, it is usually the case that calibrator sources are left unspecified until the observing window is approximately known. The CS is responsible for ensuring that appropriate calibrator sources are specified in the SB. Each execution of an SB (which may be executed more than once) is called an Execution Block (EB). The SB (or its EB realization) consists of smaller elements – scans and subscans. The SB for an INT observation typically contains scans for pointing, flux, and band-pass calibration at the beginning of each EB followed by multiple science-target scans interleaved with gain calibration and atmospheric calibration scans (see \cite{Shimojo2017}, Fig.~4), for a schematic figure of the EB structure.  In the case of mosaic observations, scans are further broken down into subscans, one for each instance of a distinct pointing.  In parallel to INT observations, fast scan TP mapping is performed with one or more total power antennas to provide full disk maps and/or FRM maps. 

\subsection{Solar Data Calibration and Quality Assurance}

Once successfully executed, data from an observing program are entered into the ALMA data archive. A copy of the ALMA data archive is available at each ARC. The data of a given PI are calibrated and assessed by DAs at the relevant ARC. Many non-solar observations are sufficiently standard that they can be automatically processed (pipeline processing). Solar observations are non-standard and are manually calibrated using specialized scripts and utilities.

Briefly, data calibration entails solving for the instrumental (complex) gain of each antenna and applying it to source data in order to infer the ``true" amplitudes and phases of the complex visibilities. This is typically done by observing a gain calibrator, usually a point-like source or a source of known structure, with a precisely known location and flux density. Taking a point-like source as an example, the source amplitude of each baseline should be that of the source itself and the phase should be zero (Fourier transform of a $\delta$-function). For a given frequency, time, and polarization only N complex gain solutions are needed whereas N(N-1)/2 observations are available. The system is highly over-determined for a large-N array like ALMA and the gains are determined using a non-linear least-squares approach. In general, the flux density of the gain calibrator is not known {\sl a priori} and it must be referenced to an observation of a flux calibrator that has a well-established and stable flux density. This is not possible for the Sun. For non-solar observations the source is too weak to contribute to system noise and the {\sl system temperature}, $T_{sys}$, is dominated by receiver noise, spillover, etc.  However, when observing the Sun the source itself contributes the bulk of the system noise, referred to as the {\sl antenna temperature,} $T_{ant}$. In order to calibrate the flux scale of solar visibility data it is necessary to measure $T_{ant}$. Therefore, while observations of a gain calibrator provide phase solutions, the amplitude scale must be established using the ALMA Calibration Device at periodic intervals throughout a solar observations. Further details are available in  \citet{Shimojo2017}.

Additional calibrations are necessary: at the beginning of an observation an antenna pointing calibration is performed, delay calibration, and the relative sideband gains. In addition, as noted in \S1, ALMA observes in four spectral windows, each 2~GHz in bandwidth and channelized into 128 frequency channels. A bandpass calibrator is therefore also observed at the beginning of an EB to deduce the relative variation of amplitude and phase across each spectral window so that it can be removed before averaging frequencies to form pseudo-continuum datasets. Additional calibrations, e.g., baseline calibration, are performed by science operations as needed.

After data calibration data analysts image each spectral window and polarization of each EB to: (i) Ensure that the data are of sufficient quality to meet the science requirements defined in the proposal (Quality Assurance level 2 – QA2), and (ii) Serve as a starting point for more advanced data analysis; e.g., time-domain imaging, self-calibration, feathering, etc. The QA2 process involves – in addition to data selection (flagging), calibration, basic imaging and image inspection -- final consistency checks of the entire dataset. QA2 has sometimes revealed issues with the approach to data reduction or have fed back into telescope operations. For example, as described in \S2.6 it was determined that the procedure used to set reference coordinates was inaccurate for data acquired in Cycle 4 and Cycle 5. The reduction script used by the DAs was therefore revised and updated to correct the issue. More recently, it was determined that, for single pointing data the reference images used standard gridding as the default which resulted in the use of the 7~m primary beam rather than the appropriately weighted primary beam response of the heterogeneous array. In any case, caution should be exercised when using the reference images for scientific analysis.

%The procedure is performed at ARCs, the PIs receive already produced images and scripts that enable reproducing entire data reduction procedure.

\section{ALMA Solar Imaging and Data Analysis}

%[I need to reference various papers from which collective experience has been drawn. Please send me anything relevant]

Current ALMA solar observing capabilities provide a great deal of flexibility in terms of frequency band selection, angular resolution, time resolution, and image size. As described in the previous section, data calibration is handled by the ARCs and is generally robust. Observers must then confront myriad issues related to solar imaging and data analysis. The Sun is arguably the most difficult imaging problem possible for ALMA because while interferometers are at their best with compact discrete sources, the Sun fills any field of view with relatively-low-contrast structure on all spatial scales. Furthermore, it is dynamic on a range of time scales and solar observers generally cannot take advantage of Earth rotation aperture synthesis to help fill in spatial scales missing from any snapshot observation as those of sidereal sources can. A major challenge is to produce high quality images that recover the critical angular scales needed to address the science objectives of interest. In this section we summarize some of the challenges and subtleties associated with imaging and analyzing ALMA solar observations. They include the use of the heterogeneous array, data weights, self-calibration, and combining single dish total power data with interferometric data.

\subsection{Heterogeneous Array}

As summarized in Section 4, solar observations with ALMA are performed by the 12-m array in combination with the ACA; that is, as a heterogeneous array comprising both 12~m and 7~m antennas. The intent is to maximize the range of spatial scales to which the observations are sensitive. On the other hand, the use of different antenna sizes, with different fields of view and weighting (\S4.3), introduces subtleties that are not yet fully understood. The effective use of the heterogeneous array is still an area of active investigation. 

\begin{figure}[b!]
\begin{center}
\includegraphics[trim={2cm 24cm 2cm 15cm},clip,width=15cm]{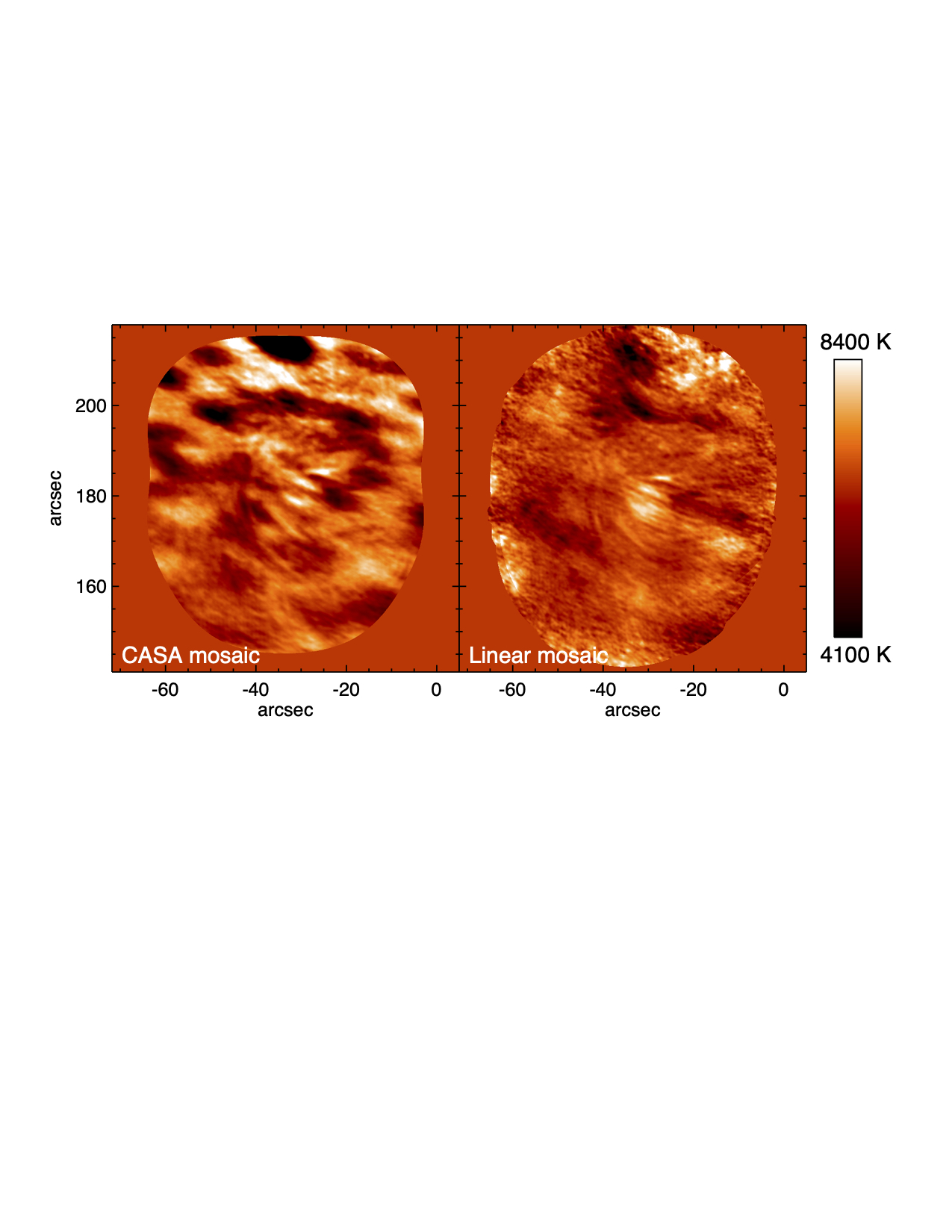}
\end{center}
\caption{Example of a small mosaic formed from ten discrete pointings in 80~s. Left panel: The image produced by joint deconvolution of the ten pointings in CASA. Right panel: The image produced by imaging each pointing separately and combining them linearly. }\label{fig:7}
\end{figure}

For a single pointing, the use of the heterogeneous array is straightforward but there are subtleties. The use of the appropriate gridding function (``mosaic" or ``mosaicft") is necessary in order to handle visibility gridding for the the different antenna sizes correctly. For sources on the solar disk, emission fills the primary beam (and beyond) for all telescopes. The issue of primary beam correction becomes somewhat complicated because the 7~m and 12~m telescopes have primary beams of different sizes, and therefore do not “see” the same region of the Sun.  In the case of a single-pointing observation, the standard procedure is to produce a map from the visibilities, deconvolve the result, then apply a primary beam correction to the deconvolved image, but neither the 7~m nor the 12~m primary beam, nor any hybrid, truly represents the appropriate correction. The correct imaging path for this case may be to have a model of the sky that is then convolved with the spatial response of each pair of telescopes, using the primary beams appropriate to each telescope, to form visibilities, and the model is then iteratively corrected to make the modeled visibilities, or the corresponding map, match the telescope data. Such a process is not currently available in CASA, but may be needed to optimally exploit ALMA solar data. 

Mosaicking works well in many cases but there appear to be failure modes that are not fully understood. The problem has been recently explored by \citet{2022FrASS...9.8115D}. Fig.~7 shows a deconvolution of a small mosaic produced by CASA, using CASA's ``mosaic'' gridding option, with an alternate approach in which the individual pointings are each mapped separately and then combined in the image plane using a linear mosaic technique and an appropriate primary beam. For this dataset most CASA images of an 80-second loop of 10-pointing mosaics have either bright or dark features that corrupt the map, and thus one cannot reliably produce time-resolved mosaics; the alternate technique, however, works reliably. The reason for the failure of CASA imaging may be due to the fact that the mosaic gridding technique limits the field of view to within a certain distance of the pointing center, but, e.g., the field of view of the 7~m antennas is larger than the field of view of the 12m antennas, and thus there is emission outside the 12~m gridded area that is sampled by the data, and is aliased into the gridded region, but this needs to be investigated further. 

Some users may find that their goals are adequately addressed by using the 12-m array alone, which greatly simplifies the data reduction and avoids some of the pitfalls described above, which are not fully understood. Others may wish to use the heterogeneous array as two homogeneous arrays: the 12-m array together with the ACA. It is straightforward to feather images made with the 12-m array, the ACA, and a TP map, again avoiding some of the issues raised by using mixed antenna pairs. 

%In the absence of such a process, we assume that the 12m primary beam is the most appropriate representation of the response of the interferometer, since 12~m dishes provide most of the collecting area and most of the visibilities, For sources within the half-power region of the 12~m primary beam, this should be satisfactory for analysis.

\subsection{Self-calibration}

Self-calibration is a technique (e.g., \citet{Cornwell1989}) where the source itself is used to determine corrections to antenna-based gains. Gain errors are most often due to atmospheric “seeing” and, as such, are dominated by phase errors. As discussed in section 2.2, ALMA makes use of WVR measurements as a proxy for phase fluctuations at each antenna for non-solar observations. This capability is not available for solar observations because the WVRs saturate when pointing at the Sun. An active topic of research is to develop an alternative proxy for phase fluctuations, such as observations of the Sun in the wing of the 183 GHz water absorption line in Band 5 (see Section 5.5). Solar observations must therefore rely on self-calibration techniques in the interim.

Experience to date suggests that a hierarchical approach to self-calibration is effective, where the average image of a scan is used as the initial source model to deduce average phase corrections. The corrected data are imaged and deconvolved to produce a new source model. This is used to determine phase corrections on a shorter time scale. This can proceed iteratively until corrections on time scales as short as the integration time are deduced. 

The self-calibration of mosaic data is somewhat less straightforward but the same hierarchical approach can work well. The difference is that it is often useful to perform the first iteration of self-calibration by determining average corrections for an entire mosaic before determining solutions for each pointing separately, using the mosaic as the model.    

\subsection{Data Weights}

Data weights enter into the imaging problem in two ways: i) the weight assigned to each visibility in forming an image through Fourier inversion, and ii) the weight assigned to each antenna beam in a mosaic. The weight assigned to a visibility measured by antenna $i$ and $j$ is $1/\sigma_{ij}^2$, where $\sigma_{ij}^2$ is the statistical variance of the measurement. For solar observations, the system noise is dominated by the source (the Sun) and the variance is of order $\sigma_{ij}\sim ({T_{ant,i}T_{ant,j}/A_i A_j})({k_B}/{\Delta\nu\Delta t})$ where $T_{ant,i}$ is the antenna temperature of antenna $i$ and $A_i$ is its effective area. All other things being equal, visibilities in a homogeneous array are given approximately the same weight (e.g., the 12-m array). However, for a heterogeneous array, we have visibility measurement with 7 m x 7m antennas, 7 m x 12 m antennas, and 12 m x 12 m antennas. The weights for each of these pairings therefore scale as $(7/12)^4:(7/12)^2:1$, or roughly 1/9:1/3:1. For a heterogeneous array of, say, 10 ACA antennas and 45 antennas from the 12-m array, the total number of baselines would be 1485. Of these, only 45 would be 7 m x 7 m pairs ($3\%$) and 450 would be 7 m x 12 m pairs ($30\%$). Coupled with the visibility weights, 7 m x 7 m correlations have very little impact on imaging and even the (weighted) 7 m x 12 m correlations come in at the $\sim10\%$ level. 

In the case of mosaic imaging, user should be aware of a second subtlety, namely, the so-called primary beam correction. 
%For a single-pointing image, the field of view is multiplied by the antenna response function, or primary beam. It is often well-described by a Gaussian and the FOV corresponds to the FWHM width of the primary beam. The final stage of imaging a field is to divide by the normalized primary beam to remove the effects of the primary beam taper. Users often correct their maps to the half-width of the primary beam but it may be desirable to extend the correction to even greater widths – the 0.35 point, for example. Users should be aware that while the signal across the FOV is corrected for the PB taper, the noise and residual artefacts are amplified with increasing angular distance from the beam center. 
The primary beam correction is formed from the grid of pointings on the sky as $\Sigma_k B(\theta-\theta_k)/\sigma_k^2$, where $B$ is the primary beam response. Consider a mosaic formed with a homogeneous array. The weight assigned to the antenna primary beam for each pointing $k$ is approximately the same for all antennas. Since the variance depends on $T_{ant}$, and $T_{ant}$ may change significantly with $k$, the weight assigned to each pointing may also vary. An extreme example is given by a Band 7 mosaic on the solar limb. As noted in Section 1.1, Band 7 uses nominal receiver settings and $T_{sys}$ is consequently small (~200 K). When pointing on the solar disk the system noise is therefore dominated by $T_{ant}$, the variance is high, and the weight assigned to that pointing is low. When the pointing is such that the limb only partially fills the beam, or the pointing is above the limb and the Sun contributes little emission to the system, the variance is low and the corresponding weight is high. This can lead to a distorted mosaic beam where the emission of interest is down-weighted relative to pointings above the limb! Users may therefore wish to correct their mosaic with a uniform weight applied to each pointing or even use weights proportional to the fractional filling factor of the Sun, although the latter idea has not been tested. 

A minor issue is that if a heterogeneous array is used, the effective primary beam (FOV) differs from that of a single antenna in a homogeneous array. The response of a single baseline with identical antennas is the same as that of a single antenna, while the response of a 7 m x 12 m baseline is geometric mean of the two antenna power patterns. The effective FOV is the weighted sum of the relevant FOV for each baseline. The net effect is to slightly broaden the 12-m array FOV ($<10\%$). 

\subsection{Combining Interferometric and Single Dish Data}

An interferometer cannot measure angular scales greater than that measured by the minimum antenna baseline. In the absence of total power information, also known as the “zero-spacing flux”, ALMA measures the distribution of brightness relative to the mean. The mean brightness of the background Sun is of order 6000-8000 K, far in excess of the low-contrast variation against the mean. The restoration of total power information to ALMA images is critical if absolute brightness temperatures are important to the science objectives. 

To zeroeth order, simply adding a constant brightness offset to an image, corresponding to the mean brightness, largely corrects for the absence of short spatial frequencies and this approach may indeed be sufficient for single-pointing imaging on the solar disk. For mosaic images, however, this is not necessarily sufficient. Solar observations with ALMA provide low-resolution full-disk maps of the Sun that are observed contemporaneously with interferometric observations using one or more of the 4 TP antennas. Calibrated TP maps contain the correct flux on all scales ranging from the 12 m primary beam size to the scale of the Sun in any given frequency band. Hence, by combining TP observations with interferometric data, one can recover the largest angular scales.

While it is possible, in principle, to first perform data combination in the uv domain followed by image deconvolution, it is perhaps more straightforward to reverse the order and to use a technique called “feathering” (e.g., Cotton 2015). Briefly, feathering begins with TP and interferometric maps, MTP and MINT, that embody low angular frequencies and high angular frequencies, respectively. The TP map must be coaligned with the interferometric map. It is convenient to cutout an imaging domain of the TP map that is perhaps twice the size of the INT map. The two are then Fourier transformed, $FT(MTP)$ and $FT(MINT)$, as is the 12 m primary beam FT(B), taken to be Gaussian. A mask is then formed from FT(B) as $m=1-FT(B)$ and the sum $FT(MTP)+mFT(MINT)$ is formed. The use of the mask is to smoothly weight down angular scales measured by the interferometer that overlap with those measured by the single dish TP map. The sum is then inverse-Fourier transformed $FT^{-1}(FT(MTP)+mFT(MINT))$, yielding the combined image. 

This approach works well in many cases; e.g., mosaics on the solar disk. However, it has shortcomings when imaging the solar limb. Two factors come into play: first, fast-scan total power maps sometimes have artefacts at the limb due to residual scan timing errors. It is undesirable to propagate these into the combined image. Second, as seen from Table 3, the minimum angular scales measured by the 12-m array decrease with frequency band and array size. While the heterogeneous array mitigates this to some extent, the degree of overlap between the TP data and interferometric data in the {\sl uv} domain, and the density of interferometric sampling in the region of overlap, both decrease with increasing frequency and array size. As a result, angular scales between the 12-m array MRS and the 12 m FOV are increasingly poorly sampled.

% \citet{Bastian2022} addressed this issue by fitting a model limb to the TP data and then used a model disk to recover angular scales more uniformly. 

\section{Future Capabilities and Observing Modes}

ALMA capabilities currently support a wide range of solar science but additional capabilities will no doubt further enrich the scientific possibilities. We briefly summarize several new instrumental capabilities that are under active consideration and then briefly touch on possible observing modes that do not depend on the development technical or data reduction capabilities but, rather, depend on observatory policy and science operations. 

\subsection{Polarimetry}

ALMA currently supports full polarimetry for non-solar observing. A longstanding goal has been to implement support of polarimetry for solar observations. Efforts are currently underway to enable polarimetry in Band 3. Polarimetry involves measuring cross-hand as well as parallel-hand correlations on each antenna baseline $ij$: i.e., $X_iX_j$ , $Y_iY_j$ , $X_iY_j$ , and $Y_iX_j$ . Formation of the Stokes polarization parameters requires measuring the phase difference between the $X$ and $Y$ polarization channels and measuring instrumental polarization embodied by “polarization leakage” terms for each antenna. The challenge has been to demonstrate that the necessary quantities can be measured using the MD mode with sufficient accuracy to ensure that the polarization signals are robust. 

The Stokes V parameter is of primary interest. It offers a diagnostic of the chromospheric magnetic field in active regions and flares. Modeling suggests \citep{Loukitcheva2017,2020FrASS...7...45L} that 3 mm may be the most promising wavelength to focus such efforts, at least initially, as it produces the highest degree of polarization of available frequency bands. Hence, initial efforts have gone into testing polarization observation in Band 3 with the intent of making Band 3 solar polarimetry a supported mode. However, testing is ongoing at the time of this writing and it is not yet known when the mode will become available. 

\begin{figure}[h!]
\begin{center}
\includegraphics[trim={2cm 4cm 2cm 4cm},clip,width=15cm]{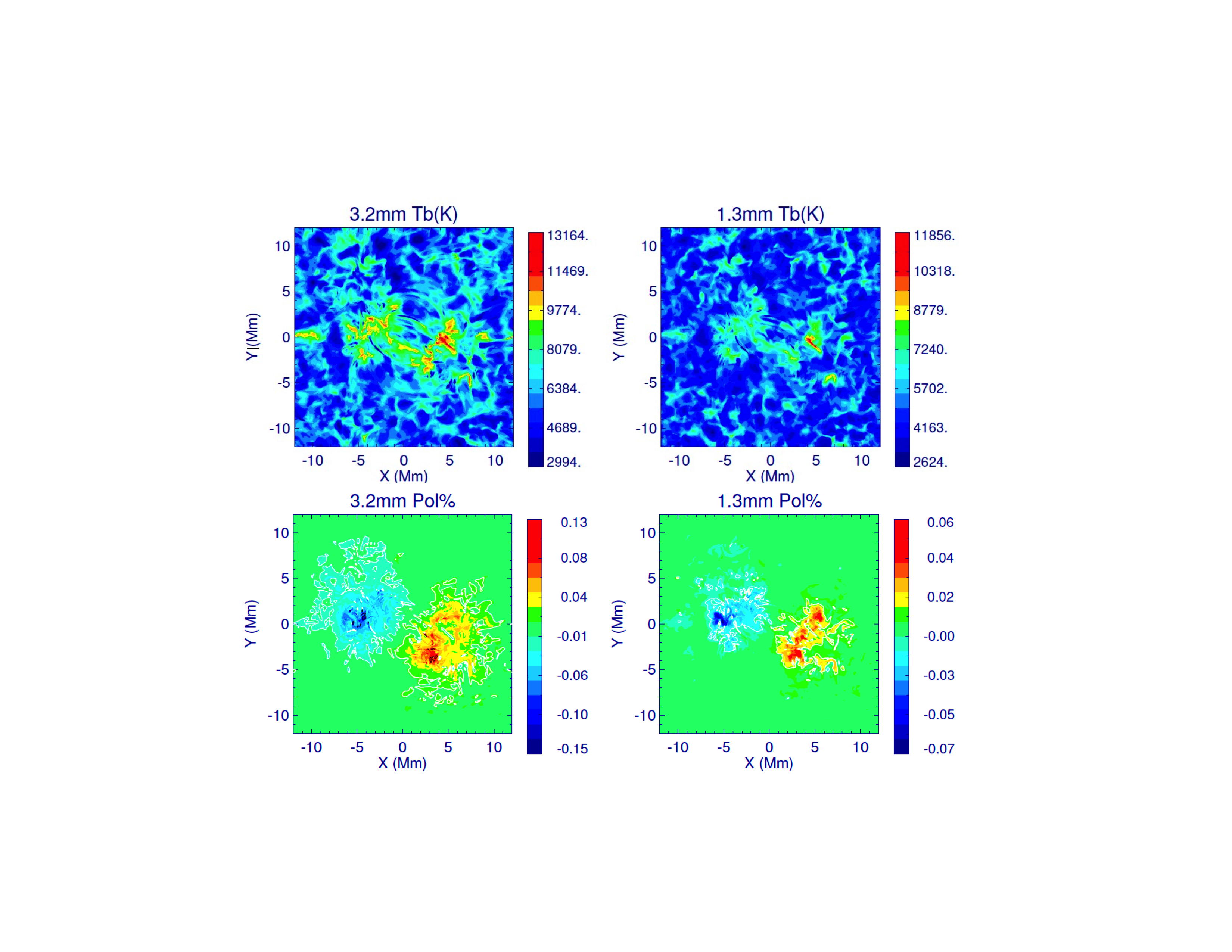}
\end{center}
\caption{Model brightness temperature distributions for a bipolar regions in Band 3 and Band 6 (top) and the corresponding degree of circular polarization (bottom). After \citet{Loukitcheva2017}.}\label{fig:8}
\end{figure}

\subsection{Spectral Line Observing}

Spectral line detections would significantly widen the diagnostic value of ALMA observations for studying the chromosphere, since they would permit line-of-sight velocity information that is missing from continuum observations. As reviewed by \citet{Wedemeyer2016}, while such lines have been observed at high millimeter frequencies, there are significant potential issues.

The two best candidates for line detection are hydrogen recombination lines (e.g., H26$\alpha$ at 353.6 GHz) and CO rotational transitions
(e.g., CO 3-2 at 345 GHz). The difficulty in detecting such lines is the fact that they are strongly pressure-broadened in the chromosphere, which reduces their contrast with the continuum. Thus, recombination lines were detected in emission by \citet{Clark2000a, Clark2000} in single-dish observations  at 662.4 (H21$\alpha$) and 888.0 (H19$\alpha$) GHz, but only at and above the solar limb where pressure is
lower, not against the solar disk. An additional complication in interpreting the spectra at these frequencies is the presence of numerous terrestrial  atmospheric absorption lines. ALMA interferometer data  should improve on these results since the relatively large beam of the single-dish maps ($\sim\!19"$) requires averaging over a large area of extended limb, resulting in a line peak 10\% above continuum level, whereas ALMA can achieve sub-arcsecond resolution (assuming successful treatment of the sharp solar limb, as discussed above). Clark et al. (2000b) argued that the line width  of the core of the recombination lines from above the limb seemed to scale roughly with frequency,  from about 550 MHz at H21$\alpha$ to 700 MHz at H19$\alpha$. If this scaling continued down to ALMA Band 7, we might expect a width of order
300 MHz for H26$\alpha$, but we note that the H21$\alpha$/H19$\alpha$ values were a factor of about 3 narrower than expected from theory 
\citep{HoangBinh1987}. As discussed earlier, ALMA can center a single baseband 2 GHz wide on a spectral line, with a frequency resolution of 16 MHz in standard solar observing mode, which would be adequate to measure a  line 300 MHz wide, but perhaps not a line that is several times wider with reduced contrast to the continuum.

\begin{figure}[b!]
\begin{center}
\includegraphics[trim={2cm 10cm 2cm 6.5cm},clip,width=18cm]{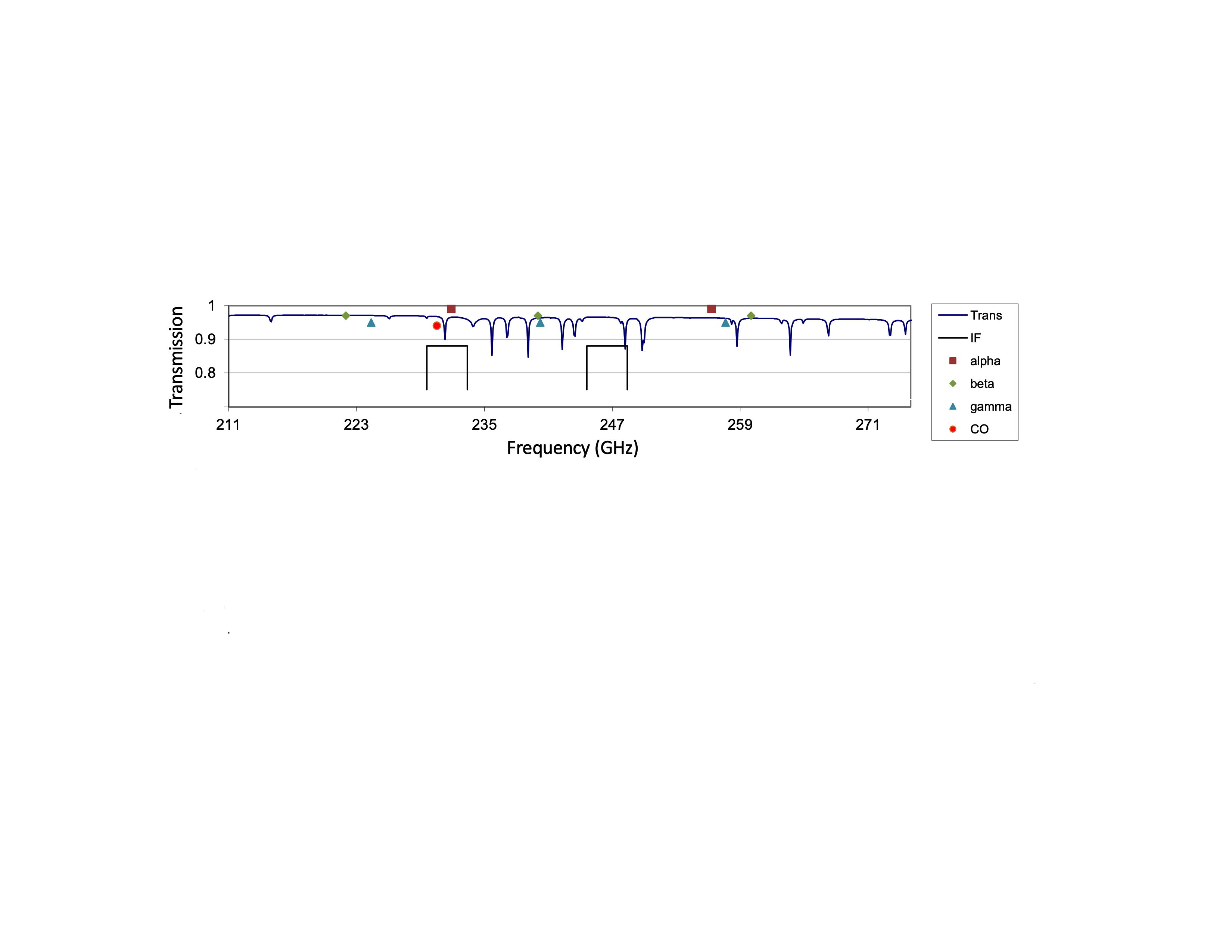}
\end{center}
\caption{Lines present in the ALMA Band 6 frequency range. The blue line indicates the atmospheric transmission and the black brackets indicate the current locations of the upper and lower sidebands for solar observing. The legend indicates the frequencies of the H$\alpha$, H$\beta$, and H$\gamma$ radio recombination lines as well as CO. Note that the lower sideband was selected to include H30$\alpha$ recombination line and the CO J=2-1 line. }\label{fig:9}
\end{figure}

There are, to this point, no confirmed detections of CO lines on the Sun at mm/submm$-\lambda$. The detections have been made so far mostly in the infrared (e.g., \citet{Ayres2006}) where a dense concentration of easily-observable CO rovibrational absorption lines occurs. Conventionally CO molecule formation requires temperatures of at most 3500 K, so one might expect CO to be preferentially detected in cool regions such as sunspot umbrae, but in fact CO can regularly be detected off-limb at heights where temperatures are expected to be much higher. An interpretation of this result is that CO forms in cool pockets such as the rarefaction phases of large-amplitude acoustic waves propagating through the chromosphere, or in the wake of chromospheric shocks. These interpretations carry the  implication that CO formation is highly dynamic. ALMA observations can help to test these models, but this will require time-resolved spectral-line mapping in order to track the dynamics of CO features. CO lines are also expected to be pressure-broadened, and may form lower in the chromosphere than the recombination lines, where pressure is higher. However, the higher mass of the CO molecule should reduce the broadening,
as found in simulations reported by Wedemeyer et al (2016). CO lines may be detected in absorption or emission, depending on the temperature profile in the region where they are found.

\subsection{Additional Frequency Bands}

Solar observing with ALMA is currently limited to four frequency bands. The use of additional frequency bands is highly desirable but requires significant testing and commissioning time. Two bands of interest are Band 1 (35-50 GHz; 7 mm) and Band 4 (125-163 GHz; 2.1 mm). Band 1 \citep{Huang2016} is an entirely new capability, not just for the solar community but for the astronomical community at large. First light with Band 1 receivers was achieved in August 2021. The goal is to make Band 1 available to the wider community in Cycle 10 (October 2023) but use by the solar community must await solar testing and commissioning. While Band 4 is part of the original suite of ALMA receivers, it has not been available for solar use. It would fill the gap between Band 3 and Band 5. Extending coverage to Band 8 (385-500 GHz; 0.7~mm) is also under consideration. 

Bands 9 (602-720 GHz), and 10 (787-950 GHz) are also of considerable interest. They are the most promising frequency bands for spectral line work. However, their use is problematic for solar INT observing. First, exceptional weather conditions are required for high-frequency observing under any circumstances. Daytime observing in Bands 9 and 10 is currently prohibited. Second, Bands 9 and 10 are double-sideband receivers and so the observing strategy may be more complicated than the MD mode used currently. On the other hand, the system temperatures for these bands is higher than for the lower frequency receivers (ranging from a few $\times\!100$~K for Band 8 to $>1000$~K for Bands 9 and 10) which may simplify their use. Both policy and technical challenges would need to be overcome to exploit them for interferometric solar observations. We note, however,  that full disk TP mapping in Band 9 was demonstrated (see \citep{Bastian2018}, Fig. 3) during commissioning and science verification in 2015 through a sky opacity $\tau\sim 1$! 

\subsection{Flare Mode Observations}

Observations of flares at mm-$\lambda$ and submm-$\lambda$ offers the potential of fundamentally new insights into energy release, particle acceleration, and emission mechanism on the Sun (see Fleishman et al, this Frontier collection). Implementation of flare mode observing confronts us with additional challenges, however. While the MD mode has successfully enabled observations of quiescent solar phenomena, it is not designed to handle the much higher flux densities produced by solar flares. Alternatives must be explored. 

It is likely that solar filters installed on the ALMA Calibration Device on each antenna will be at least part of the approach. The solar filters were the initial solution adopted by ALMA to manage solar signals. However, as detailed by \citet{Shimojo2017} the filters have a number of undesirable properties that led to the development of MD mode observing instead. The solar filter is placed in front of a given receiver in order to attenuate the incident signal. The nominal amount of attenuation is 4+2$\lambda_{mm}$ dB, amounting to signal reductions by factors of 10, 5, 3.5, and 3 for bands 3, 5, 6, and 7, respectively\footnote{A preliminary assessment of the solar filters for Band 7 showed that the attenuation introduced by the filters did not necessarily conform to expectations. In addition, single-dish TP maps made through the solar filters show poorer contrast than maps made without the filter. A painstaking assessment of the filters lies ahead.}. 

The antenna temperature is $T_{ant} = SA_e/2k_B \approx 124 S_{SFU}$~K where $S_{SFU}$ is the flux density in solar flux units ($10^{-19}$ ergs s$^{-1}$ cm$^{-2}$ Hz$^{-1}$), $A_e=\eta A$ is the effective area of the antenna, and $k_B$ is Boltzmann’s constant. For the non-flaring Sun, $T_{ant} \sim T_B \sim 6000-8000$ K. With the signal attenuation factors as given, ALMA could accommodate flares of order 600, 250, 160, and 120 SFU within the antenna beam for bands 3, 5, 6, and 7, respectively. This may be sufficient to observe many flares but may be problematic for some: the so-called “sub-THz” component of certain flares displays an inverted spectrum \citep{Krucker2013}. That is, the flux density increases with frequency and the system may saturate in the higher frequency bands.  Preliminary tests of the filters in Band 7 indicate that the degree of attenuation is less than nominal, possibly exacerbating the problem.  Alternative, or additional, strategies may be necessary to observe flares. Possibilities include the use of MD mode in tandem with solar filters and/or off-pointing the antennas so that a given flare occurs in a sidelobe of the antenna response. The first sidelobe of the 12-m antennas is a factor of 100 less sensitive than the center of the main response lobe. It is expected that testing and commissioning efforts for solar mode flare observations will begin in the coming year. 

\subsection{High Angular Resolution}

Solar observations are currently restricted to the four most compact array configurations for reasons given in \S2.2.  This limits the available angular resolution to $\approx0.6"-0.9"$. On the other hand, the high brightness of solar emission can also yield opportunities that are impossible for ``standard" interferometry of relatively weak targets for which the implicit assumption is that we need rather long observing (total signal integration) times to reach a sufficiently high signal to noise ratio (SNR). Since the image integration in ``standard” interferometry is typically performed over periods greatly exceeding the time-scale of the phase variations, the resulting image is blurred and the resolution (long-duration coherence) is lost unless provisions are made -- the use of the WVRs and self-calibration.  Phase variations occur on time scales of seconds to tens of second. Depending on the details of seeing conditions and the array configuration, these may largely manifest as low order distortions of the snapshot solar images: e.g., image wander and warping. These distortions can be, to a large extent, distinguished and separated from the true dynamics of the Sun: the large-scale solar structures change at much longer timescale (minutes) than the integration time or the time scale of phase variations over the array. The Sun allows a good SNR to be achieved even on the shortest integration times (currently 1~s, but as short as 0.16~s in principle), which enable time series of relatively sharp, but distorted images, to be obtained instead of a single blurry image by, in effect, freezing the phase variations. Self-calibration techniques are a well-established and tested means of mitigating phase variations in solar data, even in the absence of WVR corrections. It is anticipated that higher angular resolution imaging with ALMA is possible by extending observations to larger array configurations. Higher angular resolution imaging is the subject an ESO ALMA Development Study that started in July 2022.

\subsection{Standalone Total Power Mapping}

ALMA science operations currently provides TP full-disk and/or FRM maps as a complement to the interferometric observations. The full disk TP maps have been exploited scientifically in their own right \citep{Brajsa2018, Sudar2019, Selhorst2019, Alissandrakis2017, Alissandrakis2020, Alissandrakis2022}. There is interest in the solar community of using TP antennas in a mode that is independent of the 12-m array or ACA. A particularly attractive possibility is to perform multi-band TP or FRM mapping in standalone mode, possibly even in bands not currently supported for INT observing. As noted above, full disk TP mapping in Band 9 has been successfully demonstrated. Support of such a mode would require a policy change by the JAO as well as changes to the OT and operations. 

\subsection{Science Sub-arrays}

It is currently not possible to observe the Sun in more than one frequency band simultaneously. One might suppose that time sharing between two or mode frequency bands might be a promising strategy but the time overheads in switching frequency bands is prohibitive and significant changes would be needed to SBs. A more feasible approach may be the use of antenna subarrays, where the antennas of the 12-m array are distributed among two or more groups of antennas that operate as independent arrays. A user could allocate half of the antennas to one subarray and half to another; or divide them into three arrays with comparable number of antennas. However, since the number of antenna baselines and, therefore, {\sl uv} samples is $\sim\!N^2/2$ the {\sl uv} coverage provided by a given subarray quickly degrades with the number of subarrays and so users would need to consider carefully the advantages and disadvantages of the approach. 

\section{Concluding Remarks}

ALMA is a remarkable instrument that has opened a new wavelength regime for exploration. Solar observations with ALMA have been possible since late-2016 but the number of programs that have been successfully executed has been relatively modest. There are several reasons for this: i) as a general purpose instrument used by the entire astronomy and astrophysics community, competition for observing time is fierce -- only a few solar observing programs are observed per cycle; ii) as an interferometer, ALMA is not necessarily familiar to segments of the solar community -- a learning curve must be surmounted through education and outreach; iii) the data can be challenging to reduce and analyze -- progress on establishing and sharing ``best practices" is ongoing. 

Nevertheless, as the solar community becomes increasingly knowledgable about solar observing at mm/submm-$\lambda$ and as new observing modes and capabilities continue to be developed, the scientific impact of ALMA will continue to increase. This is particularly true in light of next-generation instruments coming online at other wavelengths; e.g., the Daniel K. Inouye Solar Telescope (DKIST), operating at O/IR wavelengths. Powerful synergies are available that will only increase the impact of ALMA. 

\section*{Conflict of Interest Statement}
%All financial, commercial or other relationships that might be perceived by the academic community as representing a potential conflict of interest must be disclosed. If no such relationship exists, authors will be asked to confirm the following statement: 

The authors declare that the research was conducted in the absence of any commercial or financial relationships that could be construed as a potential conflict of interest.

\section*{Author Contributions}

TB authored the bulk of the manuscript. MB wrote \S3, contributed to \S2.6. MS contributed \S2.5 and \S5.1 as well as numerous comments and corrections. SW contributed to \S2.6, and wrote \S4.1 and \S5.2.  All authors discussed and revised the manuscript and contributed numerous comments and corrections. 

\section*{Funding}
MB acknowledges support by projects  20-09922J and 21-16808J by the GACR, and project LM2015067 by the Ministry of education of the Czech Republic. SW acknowledges support from AFOSR grant 20-RV-COR-026.

\section*{Acknowledgments}

We wish to thank our friend and colleague, Prof. Richard Hills, now deceased, for both his important contributions to, and his unstinting encouragement for, making solar observations with ALMA a reality.
\smallskip

We thank the reviewers for their careful reading of the manuscript and for their constructive comments. ALMA is a partnership of ESO (representing its member states), NSF (USA) and NINS (Japan), together with NRC (Canada), MOST and ASIAA (Taiwan), and KASI (Republic of Korea), in cooperation with the Republic of Chile. The Joint ALMA Observatory is operated by ESO, AUI/NRAO and NAOJ. The National Radio Astronomy Observatory is a facility of the National Science Foundation operated under cooperative agreement by Associated Universities, Inc. This article made use of the following ALMA data: ADS/JAO.ALMA\#2011.0.00020.SV.

\bibliography{ALMA.bib}{}
\bibliographystyle{aasjournal} 
%\section*{Figure captions}

%%% Please be aware that for original research articles we only permit a combined number of 15 figures and tables, one figure with multiple subfigures will count as only one figure.
%%% Use this if adding the figures directly in the mansucript, if so, please remember to also upload the files when submitting your article
%%% There is no need for adding the file termination, as long as you indicate where the file is saved. In the examples below the files (logo1.eps and logos.eps) are in the Frontiers LaTeX folder
%%% If using *.tif files convert them to .jpg or .png
%%%  NB logo1.eps is required in the path in order to correctly compile front page header %%%

%\begin{figure}[h!]
%\begin{center}
%\includegraphics[width=10cm]{logo1}% This is a *.eps file
%\end{center}
%\caption{ Enter the caption for your figure here.  Repeat as  necessary for each of your figures}\label{fig:1}
%\end{figure}

%\begin{figure}[h!]
%\begin{center}
%\includegraphics[width=15cm]{logos}
%\end{center}
%\caption{This is a figure with sub figures, \textbf{(A)} is one logo, \textbf{(B)} is a different logo.}\label{fig:2}
%\end{figure}

%%% If you are submitting a figure with subfigures please combine these into one image file with part labels integrated.
%%% If you don't add the figures in the LaTeX files, please upload them when submitting the article.
%%% Frontiers will add the figures at the end of the provisional pdf automatically
%%% The use of LaTeX coding to draw Diagrams/Figures/Structures should be avoided. They should be external callouts including graphics.

\end{document}